\documentclass[prd,showpacs,nofootinbib,preprint]{revtex4}
\usepackage{amsmath}
\usepackage{amsfonts}
\usepackage{graphicx}
\graphicspath{{Figures/}}
\DeclareGraphicsExtensions{.pdf,.png,.jpg}

\newcommand{\be}{\begin{equation}}
\newcommand{\ee}{\end{equation}}
\newcommand{\bh}{\bar{h}}
\newcommand{\tr}{\operatorname{tr}}

\numberwithin{equation}{section}

\begin{document}

\title{Gravitational radiation from a binary system in odd-dimensional spacetime}

\author{M. Khlopunov}
\email{khlopunov.mi14@physics.msu.ru}
\affiliation{Faculty of Physics, Lomonosov Moscow State University, Moscow, 119899, Russia\\
Institute of Theoretical and Mathematical Physics, Lomonosov Moscow State University, Moscow, 119991, Russia}

\author{D.V. Gal'tsov}
\email{galtsov@phys.msu.ru}
\affiliation{Faculty of Physics, Lomonosov Moscow State University, Moscow, 119899, Russia}

\begin{abstract}
We explore possible manifestations of an odd number of extra dimensions in gravitational radiation, which are associated with violation of Huygens’ principle in flat odd-dimensional spacetime. Our setup can be regarded as the limit of an infinite compactification radius in ADD model and is not viable as realistic cosmology, but it still may be useful as a simple analytically solvable model catching certain features of more realistic scenarios. The model consists of two point masses moving inside a flat three-dimensional brane, embedded in a five-dimensional Minkowski space and interacting only through a massless scalar field localized on the same brane, while gravitational radiation is emitted into the bulk. This setup avoids the difficulties associated with taking into account the gravitational stresses binding the system, which require the cubic terms in the perturbative gravitational Lagrangian, and permits to limit ourselves to linearized theory. We calculate radiation in a linearized five-dimensional gravity generalizing the Rohrlich-Teitelboim approach to extract the emitted part of the retarded gravitational field. The source term consists of a local contribution from point particles and a non-local contribution from scalar field stresses, which is calculated using the DIRE approach to post-Newtonian expansions. In the nonrelativistic limit, we find an analog of the quadrupole formula containing an integral over the history of the particles’ motion preceding the retarded time. We also show that, for an observer on the brane, the radiation contains a third polarization: the breathing mode.
\end{abstract}

\maketitle

%%%%%%%%%%%%%%%%%%%%%%%%%%%%%%%%%%%%%%%%%%%%%%
\section{Introduction}
%%%%%%%%%%%%%%%%%%%%%%%%%%%%%%%%%%%%%%%%%%%%%%

In the last twenty years, the interest has arisen in the theories of gravitation with extra spacetime dimensions. Extra dimensions are intrinsically present in the string theory \cite{Green:1987sp}, holography \cite{Arefeva:2014kyw,Cardoso:2013vpa}, and various models proposed to solve the hierarchy problem \cite{Arkani-Hamed:1998jmv,Randall:1999ee,Randall:1999vf} and puzzles of cosmology~\cite{Dvali:2000hr,Gabadadze:2007dv,Gabadadze:2004dq,Deffayet:2000uy,Deffayet:2001pu} (for review see, e.g., \cite{Rubakov:2001kp,Maartens:2010ar,Cheng:2010pt}). New tools in astrophysics, such as gravitational-wave astronomy and black hole shadows, are opening up new ways to experimentally explore extra dimensions \cite{Visinelli:2017bny,Andriot2017,Pardo:2018ipy,Andriot:2019hay,Corman:2020pyr,Andriot:2021gwv,Vagnozzi:2019apd,Banerjee:2019nnj} as was anticipated about two decades ago \cite{Barvinsky:2003jf}. Presence of extra dimensions can manifest itself in a number of ways in gravitational-wave observations (for review see, e.g., \cite{Yu:2019jlb}). One of the most common for different theories sign of extra dimensions is the additional polarisations of the gravitational waves \cite{Andriot2017}, the experimental observation of which will become available when more gravitational-wave observatories will start operation. For compact extra dimensions, gravitational-wave signals acquire, also, the tower of massive, high-frequency Kaluza-Klein modes \cite{Andriot2017,Barvinsky:2003jf,Andriot:2019hay,Andriot:2021gwv}, which, however, are currently unavailable for observation. In some theories, gravitational waves detected by the observer on the brane are modified due to the additional contributions into the source term in the effective Einstein equations for the induced metric on the brane \cite{Shiromizu:1999wj,Maeda:2003ar,Garcia-Aspeitia:2013jea,Kinoshita:2005nx}. Also, in some theories, the falloff condition for the gravitational-wave signal is modified at the cosmological distances, compared to that for the counterpart electromagnetic signal \cite{Deffayet:2007kf}. Although, this effect was not observed in the GW170817 event \cite{Pardo:2018ipy}, it can be observed in future by LISA \cite{Corman:2020pyr}. Also, some theories predict the time gap between the observation of the gravitational-wave and electromagnetic signals from the binary neutron star merger due to the former taking the shortcut through the extra dimensions during its propagation to the observer \cite{Yu:2016tar}. The first constraints on the parameters of such theories have already been obtained from the GW170817 event data \cite{Visinelli:2017bny}.

Extra dimensions, also, modify the tidal deformabilities of the black holes and neutron stars \cite{Chakravarti:2018vlt,Cardoso:2019vof,Chakravarti:2019aup}. In particular, higher-dimensional black holes acquire non-vanishing tidal Love numbers, in contrast with the four-dimensional ones. As well, extra dimensions manifest themselves in the quasinormal modes of the black holes, which determine the spectrum of the gravitational radiation of the binary system at the final stage of its merger \cite{Chakraborty:2017qve}. Particularly, presence of extra dimensions significantly increases the damping time of the quasinormal oscillations of the black holes, providing the rigorous constraints on the parameters of corresponding theories \cite{Mishra:2021waw}. Another promising tool to probe extra dimensions is the photographs of the black holes shadows. In some theories, presence of extra dimensions results in the tidal charge of the black holes, which plays the role similar to the square of the electric charge, but can take the negative values \cite{Dadhich:2000am,Aliev:2005bi}, significantly modifying the shadows of the black holes. Parameters of the corresponding theories have already been constrained by observation of shadow of the supermassive black hole M87* \cite{Vagnozzi:2019apd,Banerjee:2019nnj,Neves:2020doc}.

For large extra dimensions one can expect manifestation of the difference between even and odd-dimensional spacetimes. Recall that in even dimensions the Green's function of the flat space d'Alembert equation is localized on the light cone, implying applicability of the standard procedure to calculate radiation. In odd dimensions Green's function has support also inside the light cone, leading to violation of the Huygens's principle and certain complications in computation of radiation. For this reason only the problems of radiation in even dimensions were considered in most of the literature \cite{Kosyakov1999,Cardoso:2002pa,Mironov:2006wi,Mironov:2007nk,Cardoso:2007uy,Kosyakov:2008wa}, while the odd dimensions were mainly discussed in the context of the radiation reaction force \cite{Galtsov:2001iv,Kazinski:2002mp,Kazinski:2005gx,Yaremko2007,Shuryak:2011tt,Dai:2013cwa,Harte:2016fru} (see, also, \cite{Kosyakov:2007qc,Kosyakov:2018wek}). Violation of Huygens's principle in odd-dimensional flat spacetime is known since the works of Hadamard \cite{hadamard2014lectures}, Courant and Hilbert \cite{courant2008methods}, Ivanenko and Sokolov \cite{Ivanenko_book}. It implies that, while in even dimensions the signal from the instant flash of the source, having reached the observer in the interval of time required for it to propagate with the speed of light, ends instantly, in odd dimensions the endless tail signal decaying with time is observed after that. On the other hand, free massless fields propagate exactly with the speed of light in all dimensions. Therefore, while the total retarded gravitational field of localized source propagates with all velocities up to that of light, its gravitational radiation, being the free field far from it, should propagate exactly with the speed of light. As a result, this mismatch makes the extraction of the emitted part of retarded gravitational field in odd dimensions a non-trivial task. In odd dimensions, there is the non-local tail contribution into the emitted part of the field analogous to that found by DeWitt and Brehme in the curved four-dimensional spacetime \cite{DeWitt:1960fc,Barack:2018yvs,Galtsov:2007zz}. However, in latter case, the tail is due to the scattering of the waves on the curvature of spacetime and its evaluation is rather complicated, while in odd dimensions the tail contribution is given in closed analytic form. The tail term can be dealt with by use of the effective field theory approach to the problems of radiation \cite{Porto:2016pyg,Cardoso:2008gn,Birnholtz:2013ffa,Birnholtz:2015hua}. However, being based on the calculations in the momentum space irrelevant to the dimensionality of the spacetime, this method does not provide us with any information about the structure of the retarded field in the wave zone and the role of the tail term in the formation of radiation. As was shown recently, this obstacle can be overcome in two ways: by Fourier transforming the retarded Green's functions over the temporal coordinate \cite{Chu:2021uea}, or by modifying the radiation definition \cite{Galtsov:2020hhn,Galtsov:2021zpb}. In this article, we follow the second approach.

To implement these ideas for the two-body system one needs a simplified model which would be tractable within the linearized gravity in Minkowski space. However, gravitational radiation in pure gravity requires second order expansion of the Ricci tensor to describe gravitational stresses binding the system. To avoid this, we propose the combination of the scalar and gravitational interactions to have the binding forces of nongravitational nature. Then gravitational radiation can be computed within the linearized gravity in Minkowski space. Our model also assumes that motion is restricted to the four-dimensional subspace (brane), while gravity operates in the odd-dimensional bulk (for simplicity we assume five dimensions). In contrary, the scalar field is assumed to live on the brane too. Then, starting with initial particles' velocities lying on the brane, we will ensure motion remains on the brane perpetually without imposing any confining mechanism. Of course, this model cannot serve a realistic model of spacetime such as ADD or RS models, and it has no purely four-dimensional interpretation. In fact, our model can be regarded as the limit of the ADD model with an infinite compactification length. This limiting theory clearly is not physically viable, but here we sacrifice viability in favor of mathematical consistency and simplicity. Our model still captures basic features of violation of Huygens's principle in odd-dimensions and demonstrate the emergence of tail and an extra polarisation in gravitational radiation for an observer living on the brane.

We use the Rohrlich-Teitelboim radiation definition \cite{Rohrlich1961,Teitelboim1970,rohrlich2007} (see, also, \cite{Kosyakov1992,Galtsov:2004uqu,Spirin2009,Galtsov:2010tny}), based on the Lorentz-invariant decomposition of the on-shell energy-momentum tensor of the retarded field, to extract the emitted part of the five-dimensional gravitational field. For consistency, we, also, explicitly take into account the contribution from the stresses of the field, through which particles forming binary system interact, into the gravitational radiation of the system. The stresses contribution is dealt with by use of the DIRE approach to the post-Newtonian expansions in the four-dimensional GR \cite{Pati:2000vt}. We find that the power of the five-dimensional gravitational radiation from the binary system is given by the analog of the famous quadrupole formula \cite{maggiore2008}, which, however, now depends on the entire history of the system's motion preceding the retarded moment of time. In contrast, in four-dimensional theory power of the gravitational radiation at some moment of time is completely determined by the state of the source at the corresponding retarded time. The obtained result is in agreement with the one found by use of the Fourier transformations of the retarded Green's functions \cite{Chu:2021uea}.

This paper is organised as follows. In the second section, we discuss the model under the consideration, polarisations of the gravitational radiation generated by the system and briefly recall the Rohrlich-Teitelboim radiation definition. Section \ref{III} is devoted to the contribution of the point particles into the gravitational radiation of the binary system. We illustrate the main steps of its computation by a simple example of the five-dimensional scalar radiation of the point particle. In Sec. \ref{IV}, we find the contribution of the scalar field's stresses, through which point particles interact with each other, into the gravitational radiation of the system by use of the DIRE approach. In section \ref{V}, we find the five-dimensional analog of the quadrupole formula for the power of the gravitational radiation of the binary system and, as for a simple example of its application, consider the gravitational radiation of the binary system on the circular orbit and corresponding orbital evolution. In Sec. \ref{VI} we discuss the obtained results and their relation to the other works.
 
%%%%%%%%%%%%%%%%%%%%%%%%%%%%%%%%%%%%%%%%%%%%%%
\section{The Setup}\label{II}
%%%%%%%%%%%%%%%%%%%%%%%%%%%%%%%%%%%%%%%%%%%%%%

As is known, the gravitational radiation of point particles interacting gravitationally differs from the electromagnetic radiation of point charges in the nonlocality of the source in the respective d'Alembert equation, which must take into account the contribution of gravitational stresses. For this reason, we cannot restrict ourselves to the linearized approximation of general relativity, but must include, as part of the radiation source, the quadratic part of the Ricci tensor as representing these stresses \cite{Weinberg:1972kfs}. To simplify this setup, we introduce a scalar field that provides the forces needed to create acceleration, assuming that this force is much greater than the force of their gravitational interaction, and the latter can be neglected. In our model, it is possible to consistently describe gravitational radiation using purely linear gravity, but still taking into account the nonlocality of the source in the d'Alembert equation for the gravitational field $h_ {MN}$ due to the contribution of interaction field stresses. The consistency of such a model will be ensured if the interacting particles do not leave the four-dimensional subspace (brane); this is guaranteed because the scalar field is restricted to live on the brane. We do not specify the physical mechanism of such localization and do not introduce brane as a physical object. Our goal is to reveal the features of gravitational radiation in odd-dimensional space-time due to violation of the Huygens principle, which induces  the effect of nonlocality of a different nature.

%%%%%%%%%%%%%%%%%%%%%%%%%%%%%%%%%%%%%%%%%%%%%%
\subsection{Brane and bulk}

The model assumes that the system of point particles
$m_a$ with initial velocities, bounded by the four-dimensional subspace of the five-dimensional spacetime, moves under mutual scalar forces, forever remaining in this subspace, which we loosely call the three-brane. We choose the five-dimensional coordinates $x^M=(x^\mu, x^4)$ with $M = \overline{0,4}$ and $\mu = \overline{0,3}$, so that the brane is orthogonal to $x^4$ and is parametrized by the same bulk coordinates. 
 
We start with the generally covariant action
\be
S = S_g + S_\varphi + S_m,
\ee
which is the sum of the five-dimensional Einstein-Hilbert term
\be
S_g = \frac{1}{2\kappa_{5}} \int d^{5}x \sqrt{-g} \,  R,
\ee
where $\kappa_{5}$ is the five-dimensional gravitational constant, the four-dimensional scalar field term
\be
S_\varphi = - \frac{1}{8\pi} \int d^{4} \sigma \sqrt{-\gamma} \gamma^{\mu\nu} \partial_{\mu} \varphi(\sigma) \partial_{\nu} \varphi(\sigma),
\ee
where the induced metric $\gamma_{\mu\nu}=g_{\mu\nu}$ coincides with the four-dimensional part of the bulk metric $g_{MN}$, chosen in the gauge $g_{4\mu}=0$, and the particle term describes their motion along affinely parameterized world lines $z_{a}^{\mu}(\tau)$ (omitting an index $a$ in $\tau_a$)
\be
S_p = - \sum_a \int \left( m_a + g_a \varphi(z_a) \right) \left(-\dot{z}_{a}^{2}\right)^{1/2}d\tau, \qquad \dot{z}_{a}^{2} = \dot{z}_{a}^{\mu} \dot{z}_{a}^{\nu} \gamma_{\mu\nu},\qquad \dot{z}_{a}^{\mu} =d z_{a}^{\mu}/d\tau.
\ee
The scalar charges $g_a$ with this definition are dimensionless.
 
The corresponding Einstein equations will read
\be
\label{eq:grav_EOM_gen}
R_{MN} - \frac{1}{2} g_{MN} R = 2 \kappa_{5} T_{MN}, \qquad  T_{MN} = \delta^{\mu}_{M} \delta^{\nu}_{N} \, T_{\mu\nu} \, \delta(x^4),
\ee
where $T_{\mu\nu}$ is the (restricted to the brane) matter energy-momentum tensor consisting of the particles and scalar field terms $T_{\mu\nu} = T_{\mu\nu}^{\rm P} + T_{\mu\nu}^{\rm F}$:  
\begin{align}
&T_{\mu\nu}^{\rm P} = \sum_a \int d\tau  \left( m_a + g_a \varphi(z_a) \right) \dot{z}_{a\mu} \dot{z}_{a\nu} \frac{\delta^{(4)}(x-z_{a})}{\sqrt{-\gamma}},\label{eq:pp_EMT_gen}\\
\label{eq:sc_EMT_gen}
&T_{\mu\nu}^{\rm F} = \frac{1}{4\pi} \left( \partial_{\mu} \varphi \partial_{\nu} \varphi - \frac{1}{2} \gamma_{\mu\nu} (\partial \varphi)^2 \right).
\end{align}
In accordance with the localisation of the scalar field on the three-brane, it contains the distributional singularity provided by the five-dimensional delta function under the four-dimensional integral over the brane.

We also find the equations of motion for the matter localised on the three-brane. So, scalar field equation of motion has the form
\be
\label{eq:sc_EOM_gen}
{^{(\gamma)}}\square \varphi = \frac{1}{\sqrt{-\gamma}} \partial_\mu \left( {\sqrt{-\gamma}} \gamma^{\mu\nu} \partial_\nu \varphi \right) = 4\pi \sum_{a} g_{a} \int d\tau \frac{\delta^{(4)}(x - z_a)}{\sqrt{-\gamma}},
\ee
Equations of motion for the point particles can be presented in the form
\be
\label{eq:pp_EOM_gen}
m_a(\tau)\frac{Dz^{\mu}_{a}}{d\tau}=g_a\left( \gamma^{\mu\nu}+z_a^\mu z_a^\nu \right)\partial_\nu\varphi,
\ee
where the variable mass is introduced
\be
m_a(\tau)=m_a + g_a\varphi.
\ee

In fact, the Einstein equations (\ref{eq:grav_EOM_gen}) with point particle sources only make sense in the linearized approximation, so our generally covariant formulation was only intended to emphasize the geometric meaning of the  quantities  under consideration. Now we assume
\be
g_{MN}  = \eta_{MN} + h_{MN} , \quad \left \vert h_{MN} \right \vert \ll 1.
\ee
with the Minkowski metric $\eta_{MN} \equiv {\rm diag} (-1,1,1,1,1)$, and in what follows perform manipulations with indices using the flat metric.
Introducing the reduced field  
\be
\label{eq:bh_def}
\bar{h}_{MN} \equiv h_{MN} - \frac{1}{2} \eta_{MN} h,\qquad h = \eta^{MN} h_{MN},
\ee
and imposing the Lorentz gauge condition
\be
\label{eq:Lorentz_gauge}
\partial^{M} \bh_{MN} = 0,
\ee
we find the linearised Einstein's equation as
\be
\label{eq:lin_Ein_eq}
{^{(5)}}\square \bh_{MN}  = - 2 \kappa_{5} T_{MN},
\ee
where ${^{(5)}}\square = \eta^{MN} \partial_M \partial_N$ is the d'Alembert operator on the five-dimensional Minkowski space. Note, for the Lorentz gauge condition \eqref{eq:Lorentz_gauge} to be valid, the matter energy-momentum tensor has to be conserved
\be
\partial^{M} T_{MN} = 0,
\ee
what is easily verified using the above definitions, and the four-dimensional conservation equation
\be
\partial^{\mu} T_{\mu\nu}  = 0,
\ee
which holds by virtue of the equations of motion
(\ref{eq:sc_EOM_gen},\ref{eq:pp_EOM_gen}).

%%%%%%%%%%%%%%%%%%%%%%%%%%%%%%%%%%%%%%%%%%%%%%%
\subsection{Gravitational radiation: polarizations}

We want to calculate the 5D gravitational radiation of our system. To do this, we introduce another division of coordinates: $x^M=(x^0=t, x^i), \, i=\overline{1,4}$.

The energy-momentum flux carried by the 5D transverse-traceless wave polarizations is given by an effective energy-momentum tensor analogous to that in \cite{maggiore2008}
\be
\label{eq:GW_EMT_gen}
t_{MN} = \frac{1}{4 \kappa_{5}} \left \langle \partial_{M} h_{ij}^{\rm TT} \partial_{N} h_{ij}^{\rm TT} \right \rangle,
\ee
where we assume that the motion of the source is periodic and the brackets $\langle \ldots \rangle$ denote the averaging over the period. In the transverse-traceless gauge we use here, the following conditions hold:  
\be
\label{eq:TT_gauge_cond}
h_{0M}^{\rm TT} = 0, \quad h_{ii}^{\rm TT} = 0, \quad \partial^{j} h_{ij}^{\rm TT} = 0,
\ee
which ensure that we deal with five polarisations. These can be obtained acting on the unconstrained potentials by the projector \cite{maggiore2008}
\begin{align}
\label{eq:TT_gauge_def}
& h_{ij}^{\rm TT}(x) = \Lambda_{ij,kl}(\mathbf{n}) h_{kl}(x), \quad \mathbf{n} = \frac{\mathbf{x}}{|\mathbf{x}|} \\
& \Lambda_{ij,kl}(\mathbf{n}) \equiv P_{ik} P_{jl} - \frac{1}{3} P_{ij} P_{kl}, \quad P_{ij}(\mathbf{n}) \equiv \delta_{ij} - n_{i} n_{j},
\end{align}
which has the vanishing traces
\be
\label{eq:Lambda_traces}
\Lambda_{ii,kl}(\mathbf{n}) = \Lambda_{ij,kk}(\mathbf{n}) = 0.
\ee
Therefore, the transverse-traseless potentials can be equivalently obtained from the trace-reversed metric deviations
\be
\label{eq:TT_gauge_practical}
h_{ij}^{\rm TT}(x) = \Lambda_{ij,kl}(\mathbf{n}) \bh_{kl}(x).
\ee
From five polarizations of the bulk modes, three are present in the brane subspace. Choosing the propagation vector $\mathbf{n}$ to be aligned with $x^{3}$-direction, we will have
\be
\bar{h}_{ij}^{\rm TT}(x) = 
\begin{pmatrix}
h_{+} - \displaystyle \frac{1}{2}h_{\displaystyle \circ} & h_{\times} & 0 & h_{14} \\
h_{\times} & - h_{+} - \displaystyle \frac{1}{2}h_{\displaystyle \circ} & 0 & h_{24} \\
0 & 0 & 0 & 0 \\
h_{14} & h_{24} & 0 & h_{\displaystyle \circ} 
\end{pmatrix},
\ee
where the "cross" and "plus" polarizations are standard  
\be
\label{eq:pl_cr_pol_part}
h_{+} = \frac{1}{2}\left( h_{11} - h_{22} \right), \quad h_{\times} = h_{12},
\ee
and the $h_{\displaystyle \circ}$ is the "breathing" polarization \cite{Andriot2017} given as
\be
\label{eq:breath_pol_part}
h_{\displaystyle \circ} = \frac{2}{3}h_{44} - \frac{1}{3}\left( h_{11} + h_{22} \right).
\ee
So the brane-living observer will see only three polarizations  -- cross, plus, and the breathing mode \cite{Andriot2017}. From the linearized geodesic deviation equation \cite{maggiore2008,Andriot2017}, it can be seen that under the action of breathing mode the circle of brane-living test masses lying in the plane orthogonal to the propagation direction will uniformly shrink and stretch.

Polarizations amplitudes for a geneneric symmetric tensor of second order $A_{ij}$ in transverse-traceless gauge for an arbitrary propagation direction $\mathbf{n}$ are given in appendix A.

%%%%%%%%%%%%%%%%%%%%%%%%%%%%%%%%%%%%%%%%%%%%%%
\subsection{Dynamics of the non-relativistic binary}

In the linear approximation \eqref{eq:lin_Ein_eq},  the particles and the scalar field propagate on the flat Minkowski background $\gamma_{\mu\nu} = \eta_{\mu\nu}$, and the equations of motion \eqref{eq:sc_EOM_gen} and \eqref{eq:pp_EOM_gen} reduce to
\begin{align}
\label{eq:sc_EOM_flat_br}
&{^{(4)}}\square \varphi (x) = 4\pi \sum_{a} g_a \int d\tau \, \delta^{(4)} \left( x - z_a \right), \\
\label{eq:pp_EOM_flat_br}
&\left( m_a + g_a \varphi(z_a) \right) \ddot{z}_{a}^{\mu} = - g_a \partial^{\mu} \varphi(z_a) - g_a \partial_{\nu} \varphi(z_a) \dot{z}_{a}^{\nu} \dot{z}_{a}^{\mu},
\end{align}
where ${^{(4)}}\square = \eta^{\mu\nu} \partial_{\mu} \partial_{\nu}$ is the d'Alembert operator on the four-dimensional Minkowski space. The particles' motion is completely governed by their mutual interaction through the scalar field given by the retarded solution of Eq. \eqref{eq:sc_EOM_flat_br}.

In what follows, we are interested in the gravitational radiation from the non-relativistic binary system, in  which case Eq. \eqref{eq:pp_EOM_flat_br} simplifies significantly. Moreover, we assume accordingly:
\begin{itemize}
\item 
the particles are non-relativistic $|\dot{\mathbf{z}}_a| \ll 1$;
\item
spatially separated $|\mathbf{z}_{2} - \mathbf{z}_{1}| \gg - g_{1}g_{2}/m_{a}, \, \forall a$;
\item
the scalar field just governs the motion of particles and is not radiated.
\end{itemize}
Then, we neglect the terms proportional to the square of particle's velocity $\sim \dot{z}_{a}\dot{z}_{a}$ and the self-interaction terms
\be
\label{eq:self-int_neglect}
\varphi(z_{a}) z_{a} = \varphi_{b}(z_{a}) z_{a}, \quad b \neq a
\ee
in the Eq. \eqref{eq:pp_EOM_flat_br}, where $\varphi_a(x)$ is the scalar field produced by $a$-th particle. So the scalar field of $a$-th particle is a Coulomb field \cite{Jackson:1998nia}:
\be
\label{eq:ret_sc_non-rel}
\varphi_{a}(x) = - \frac{g_a}{|\mathbf{x} - \mathbf{z}_{a}(t)|},
\ee
where $t=x^0$, and we neglected the retardation. Thus, combining Eqs. \eqref{eq:self-int_neglect} and \eqref{eq:ret_sc_non-rel} with the condition of spatial separation, we neglect the influence of the scalar field on particles' masses
\be
(m_a + g_a \varphi_{b}(z_a)) \simeq m_a.
\ee

As a result, we obtain the non-relativistic particles equation of motion simply as
\begin{equation}
\label{eq:pp_EOM_non-rel}
m_{a} \ddot{\mathbf{z}}_{a} = - g_{a} \nabla \varphi_{b} (\mathbf{z}_{a}),
\end{equation}
where $\varphi_b(x)$ is given by the Eq. \eqref{eq:ret_sc_non-rel}. Note that these four-dimensional equations  of motion admit the stable elliptical orbits. The point particles and scalar field energy-momentum tensors reduce to
\begin{align}
\label{eq:pp_EMT_brane}
&T_{MN}^{\rm P} = \delta_{M}^{\mu} \delta_{N}^{\nu} \sum_{a} m_a \int d\tau \, \dot{z}_{a\mu} \dot{z}_{a\nu} \, \delta^{(4)}(x - z_a) \delta(x^4),\\
\label{eq:sc_EMT_brane}
&T_{MN}^{\rm F} = \frac{1}{4\pi} \delta_{M}^{\mu} \delta_{N}^{\nu} \left(\partial_{\mu} \varphi \partial_{\nu} \varphi - \frac{1}{2} \eta_{\mu\nu} \partial^{\alpha} \varphi \partial_{\alpha} \varphi \right) \delta(x^4).
\end{align}
Clearly, as we tacitly assumed the initial particles' velocities to lie on the brane, no forces normal to the brane will arise, so they perpetually remain there. Summarising, we consider the binary system of non-relativistic, spatially separated point particles moving inside the flat three-brane and interacting with each other only through the scalar field localised on the same brane. Dynamics of the binary is described by Eq. \eqref{eq:pp_EOM_non-rel}, while the scalar fields produced by the particles is given by Eq. \eqref{eq:ret_sc_non-rel}. Gravitational waves will arise as the solution of the five-dimensional d'Alembert equation Eq. \eqref{eq:lin_Ein_eq} with the matter energy-momentum tensor given as the sum of Eqs. \eqref{eq:pp_EMT_brane} and \eqref{eq:sc_EMT_brane} which both contain delta-functions with the support on the brane. Since this sum is four-dimensionally divergenceless on-shell, it will be automatically five-dimensionally divergenceless, since the particle velocities have zero component normal to the brane, while the scalar field has no dependence on the fifth coordinate. This condition is necessary for consistency of the Eq. \eqref{eq:lin_Ein_eq}. 

%%%%%%%%%%%%%%%%%%%%%%%%%%%%%%%%%%%%%%%%%%%%%%
\subsection{Gravitational waves generated by the system}

Gravitational waves generated by the binary system are given by the retarded solution of Eq. \eqref{eq:lin_Ein_eq} constructed by the corresponding Green's function \cite{courant2008methods,hadamard2014lectures,Ivanenko_book,Galtsov:2001iv,Kazinski:2002mp,Cardoso:2002pa,Dai:2013cwa,Galtsov:2020hhn}
\begin{align}
\label{eq:GW_gen}
& \bh_{MN}(x) = - 2 \kappa_{5} \int d^{5}x' \, T_{MN}(x') \, G_{\rm ret}^{4+1} (x - x'), \\
\label{eq:5D_Green}
& G_{\rm ret}^{4+1}(x) = \frac{\theta(t)}{2 \pi^2} \left \lbrack \frac{1}{2} \frac{\theta(- x^2)}{(- x^2)^{3/2}} - \frac{\delta(- x^2)}{(- x^2)^{1/2}} \right \rbrack.
\end{align}
Note that the Green's function \eqref{eq:5D_Green} is localised not only on the light cone $x^2 = 0$, but also inside it, leading to the propagation of the retarded gravitational field in space with all velocities up to that of light. However, further we demonstrate, using the Rohrlich-Teitelboim approach to radiation, that the retarded field derivative contains the component propagating exactly with the speed of light and representing the emitted part of the field.

Also, while each term in Eq. \eqref{eq:5D_Green} is separately divergent on the light cone, the resulting field is finite \cite{Galtsov:2020hhn}. Let us briefly discuss the regularisation procedure used to make the finiteness of the field explicit. We consider each term in Eq. \eqref{eq:5D_Green} as a product of two distributions, e.g., their derivatives are given by the Leibniz rule
\begin{align}
\label{eq:distr_diff_rule_1}
& \frac{d}{dy} \left( y^{n} \delta(y) \right) = n y^{n-1} \delta(y) + y^{n} \delta'(y), \\
\label{eq:distr_diff_rule_2}
& \frac{d}{dy} \left( y^{n} \theta(y) \right) = n y^{n-1} \theta(y) + y^{n} \delta(y).
\end{align}
It is clear that if $n$ is negative, these distributions are singular at $y = 0$. We introduce the regularising parameter $\epsilon \to +0$ by shifting the arguments of delta and Heaviside functions as $y \to y + \epsilon$. Then, integrating by parts the Heaviside function term in \eqref{eq:GW_gen} we can extract the divergence contained in it and cancel it out with the divergent delta function term coming to the finite result. The same procedure is valid when one considers the derivative of retarded field, taking into account the differentiation rules \eqref{eq:distr_diff_rule_1} and \eqref{eq:distr_diff_rule_2}.

We consider the point particles and scalar field contributions into the gravitational waves separately, decomposing the retarded gravitational field \eqref{eq:GW_gen} into two terms
\be
\bh_{MN}(x) = \bh_{MN}^{\rm P}(x) + \bh_{MN}^{\rm F}(x).
\ee
Point particles contribution is computed using the Rohrlich-Teitelboim approach to radiation, while the scalar field term is calculated by use of the DIRE approach to the post-Newtonian expansions (for review see, e.g., \cite{Pati:2000vt}).

%%%%%%%%%%%%%%%%%%%%%%%%%%%%%%%%%%%%%%%%%%%%%%%
\subsection{ADD-model with the infinite compactification radius}

Let us see how  our model can be interpreted as the limiting case of the ADD-model with the infinite compactification radius of an extra dimension. We start with the five dimensional ADD-model with the compactification radius $R$. The scalar retarded Green's function of the d'Alembert equation in such a theory is written as \cite{Giudice:1998ck}
\be
G_{\rm ret}^{\rm ADD}(x;y) = - \frac{1}{2\pi R} \sum_{n=-\infty}^{+\infty} \int \frac{d^4 k}{(2\pi)^4} \frac{e^{im_{n}y}e^{ik_{\alpha}x^{\alpha}}}{k_{\mu}^2+m_{n}^2},
\ee
where $y$ is the coordinate along the extra dimension and $m_n = n/R$ are the Kaluza-Klein masses. Integrating over $k^0$ in accordance with the retardation condition we find it as
\be
G_{\rm ret}^{\rm ADD}(x;y) = - \frac{\theta(t)}{2\pi R} \sum_{n=-\infty}^{+\infty} \int \frac{d^3 k}{(2\pi)^3} \frac{\sin \sqrt{\mathbf{k}^2 + m_{n}^{2}}t}{\sqrt{\mathbf{k}^2 + m_{n}^{2}}} e^{i\mathbf{k}\mathbf{x}} e^{im_{n}y}.
\ee
Performing integration over the angular variables we obtain
\be
G_{\rm ret}^{\rm ADD}(x;y) = \frac{\theta(t)}{4 \pi^3 R r} \frac{\partial}{\partial r} \sum_{n=-\infty}^{+\infty} e^{im_{n}y} \int dk \, \cos{kr} \frac{\sin \sqrt{k^2 + m_{n}^{2}}t}{\sqrt{k^2 + m_{n}^{2}}}.
\ee
Evaluating the remaining integral \cite{Ivanenko_book,zwillinger2014table}
\be
\int dk \, \cos{kr} \frac{\sin \sqrt{k^2 + m_{n}^{2}}t}{\sqrt{k^2 + m_{n}^{2}}} = \frac{\pi}{2} J_{0}(m_{n} \sqrt{t^2 - r^2}) \left \lbrack \theta(t-r) + \theta(t+r) - 1 \right \rbrack
\ee
and differentiating over the spatial distance, we get
\begin{align}
G_{\rm ret}^{\rm ADD}(x;y) &= \frac{\theta(t)\theta(t-r)}{8 \pi^2 R} \sum_{n=-\infty}^{+\infty} e^{im_{n}y} \frac{m_n J_1 (m_n \sqrt{t^2 - r^2})}{\sqrt{t^2 - r^2}} -\nonumber\\ &-\frac{\theta(t)\delta(t-r)}{8 \pi^2 R r} \sum_{n=-\infty}^{+\infty} e^{im_{n}y} \times   \times J_0 (m_n \sqrt{t^2 - r^2}).
\end{align}
Two series over the Kaluza-Klein masses can be combined using the following relations for the products of Heaviside and delta functions
\be
\theta(t)\theta(t-r) = \theta(t)\theta(t^2-r^2), \quad \frac{\theta(t)\delta(t-r)}{r} = 2\theta(t)\delta(t^2-r^2),
\ee
arriving at the following expression for the ADD retarded Green's function
\be
G_{\rm ret}^{\rm ADD}(x;y) = - \frac{1}{\sqrt{t^2 - r^2}} \frac{\partial}{\partial \sqrt{t^2-r^2}} \left \lbrack \frac{\theta(t)\theta(t^2 - r^2)}{8\pi^2R} \sum_{n=-\infty}^{+\infty} e^{im_{n}y} J_0 (m_n \sqrt{t^2 - r^2}) \right \rbrack.
\ee
In the limit of infinite compactification radius 
$R\to\infty$, one can pass to continuous limit $m_n \to m \in \mathbb{R}$ and integrate over $m$:  
\be
G_{\rm ret}^{R\to\infty}(x;y) = - \frac{1}{\sqrt{t^2 - r^2}} \frac{\partial}{\partial \sqrt{t^2-r^2}} \left \lbrack \frac{\theta(t)\theta(t^2 - r^2)}{4\pi^2} \int_{0}^{+\infty} dm \, \cos(my) J_0 (m \sqrt{t^2 - r^2}) \right \rbrack,
\ee
where we have taken into account the positive parity of $J_0(x)$. Evaluating the resulting integral \cite{zwillinger2014table}, we arrive at the following expression for the Green's function
\be
G_{\rm ret}^{R\to\infty}(x;y) = - \frac{1}{\sqrt{t^2 - r^2}} \frac{\partial}{\partial \sqrt{t^2-r^2}} \frac{\theta(t)\theta(t^2-r^2-y^2)}{4\pi^2\sqrt{t^2-r^2-y^2}}.
\ee
Finally, after the differentiation we come to the expression equivalent to the retarded Green's function of the d'Alembert equation in the five-dimensional Minkowski space \eqref{eq:5D_Green}
\be
G_{\rm ret}^{R\to\infty}(x;y) = \frac{\theta(t)}{2\pi^2} \left \lbrack \frac{1}{2} \frac{\theta(t^2-r^2-y^2)}{(t^2-r^2-y^2)^{3/2}} - \frac{\delta(t^2-r^2-y^2)}{\sqrt{t^2-r^2-y^2}} \right \rbrack.
\ee

%%%%%%%%%%%%%%%%%%%%%%%%%%%%%%%%%%%%%%%%%%%%%%%
\subsection{Rohrlich-Teitelboim definition of radiation}

In four dimensions, the retarded electromagnetic field of the point charge consists of two terms: one proportional to $1/r^2$ and being the deformed Coulomb-like part of the field, and another, which depends on the acceleration of the charge and scales as $1/r$. The latter gives a non-zero flux of the field energy-momentum through the distant sphere, and therefore represents radiation. To argue that this is radiation indeed, Rohrlich \cite{Rohrlich1961,rohrlich2007} and Teitelboim \cite{Teitelboim1970} (see also \cite{Kosyakov1992,Galtsov:2004uqu,Kosyakov:2007qc,Galtsov:2010tny,Kosyakov:2018wek}), computed the most long-range part of the on-shell energy-momentum tensor, demonstrating that it exhibits special properties, meaning that the corresponding part of the field energy-momentum propagates at the speed of light.

The Rohrlich-Teitelboim approach is based on the use of certain covariantly defined quantities, so we briefly recall their definitions. Let us consider a point particle moving along a worldline $z^{M}(\tau)$ with the $D$-velocity $v^{M}=dz^{M}/d\tau$. The observation point coordinates are denoted as $x^{M}$. Consider the observation point as a top of the light cone in the past, and denote the intersection point of this light cone with the particle's worldline as $\hat{z}^{M}={z}^{M}(\hat{\tau})$. The quantity $\hat{\tau}$ is called the retarded proper time and is defined by the equation
\begin{equation}
\label{eq:ret_prop_time_eq}
\left( x^{M} - z^{M} (\hat{\tau}) \right)^2 = 0, \quad x^0 \geq z^0(\hat{\tau}).
\end{equation}
If the signal propagating with the speed of light were emitted at the retarded moment, it would reach the observation point at the corresponding moment of time $x^{0}$. Further, all the hatted quantities correspond to the retarded proper time $\hat{\tau}$. We introduce three spacetime vectors: a lightlike vector $\hat{X}^{M} = x^{M} - \hat{z}^{M}$ directed from $\hat{z}^{M}$ to the observation point; a spacelike unit vector $\hat{u}^{M}$, orthogonal to the particle's velocity $\hat{v}^{M}$; and a lightlike vector $\hat{c}^{M} = \hat{v}^{M} + \hat{u}^{M}$ being the sum of the formers. These vectors have the following properties:
\begin{align}
\label{eq:ret_q_1}
&\hat{v}^2 = - \hat{u}^2 = -1; \quad \hat{c}^2 = 0; \quad (\hat{c}\hat{v}) = - (\hat{c}\hat{u}) = -1; \quad (\hat{v}\hat{u}) = 0, \\
\label{eq:ret_q_2}
&\hat{X}^{M} = \hat{\rho} \hat{c}^{M}; \quad \hat{\rho} = - (\hat{v}\hat{X}), \quad \hat{X}^2 = 0,
\end{align}
where $(AB)=A^{M}B_{M}$. Note that $\hat{\rho}$ is a Lorentz-invariant distance equal to the spatial distance in the Lorentz frame comoving with the particle at the retarded moment $\hat{\tau}$. Note also that for a point charge, moving along the worldline $z^{M}(\tau)$ for an infinite proper time $-\infty<\tau<\infty$, certain care is needed to correctly define the asymptotic conditions for acceleration, for details see \cite{Teitelboim1970}. Here we will not discuss this subtlety, considering the simple case of periodic motion. Far from the region of the particle's motion $\hat{\rho} \sim r$, so the Lorentz-invariant definition of the distance is equivalent to the naive definition. Nevertheless, in order to obtain Lorentz-covariant expansions of tensors, one has to use $1/\hat{\rho} $ as an expansion parameter. 
 
Based on the definitions above, in four-dimensional electrodynamics Teitelboim \cite{Teitelboim1970} obtained the following decomposition of the the on-shell energy-momentum tensor:
\be
T^{\mu\nu} = T_{\rm Coul}^{\mu\nu} + T_{\rm mix}^{\mu\nu} + T_{\rm rad}^{\mu\nu},
\ee
where the Coulomb part falls down as $\hat{\rho}^{-4}$, the mixed part -- as $\hat{\rho}^{-3}$, and the last part -- as $\hat{\rho}^{-2}$. Teitelboim found that the most long-range term of the on-shell energy-momentum tensor has the following properties:
\begin{itemize} 
\item 
It is separately conserved $\partial_{\nu} T_{\rm rad}^{\mu\nu}=0$.
\item
It is proportional to the direct product of two null vectors $\hat{c}^\mu \hat{c}^\nu$ and, therefore, $\hat{c}_\mu T_{ \rm rad}^{\mu\nu}=0$.
\item 
It scales as $1/\hat{\rho}^2$ and gives positive definite energy-momentum flux through the distant sphere.
\end{itemize}
It is clear that the corresponding part of the field's energy-momentum propagates with the speed of light. Therefore, the radiation power can be computed as the flux of the energy, associated with $T_{\rm rad}^{\mu\nu}$. Note that, due to the electromagnetic field's energy-momentum tensor being the bilinear functional of the field derivatives $T_{\mu\nu} \sim \partial A \, \partial A$, one can define the emitted part of the latter $\left( \partial_{\mu} A_{\nu} \right)^{\rm rad}$, which contributes into the radiated part of the energy-momentum tensor $T_{\rm rad}^{\mu\nu}$.

Similar decomposition and reasoning holds in any spacetime dimensions \cite{Kosyakov1999,Kosyakov:2008wa,Spirin2009} and, also, for the scalar \cite{Spirin2009,Galtsov:2020hhn} and gravitational fields. The only difference is that in $D$ dimensions area of the distant sphere is proportional to $r^{D-2}$, thus, the relevant asymptotic behaviour of the field derivative in the wave zone is $1/r^{D/2-1}$.

Passing to our theory of the five-dimensional gravitational radiation, we calculate the long-range part (with respect to the Lorentz-invariant distance $\hat{\rho}$) of derivative of the retarded gravitational field \eqref{eq:GW_gen} and substitute it into the bilinear functional \eqref{eq:GW_EMT_gen}. All the listed properties of the emitted part of the energy-momentum tensor $T_{\rm rad}^{MN}$ hold for the obtained tensor. Note that the transformation into the transverse-traceless gauge can be performed after the extraction of the emitted part of the field derivative, because the derivative of the $\Lambda$-tensor is proportional to the inverse distance $\partial_{m} \Lambda_{ij,kl}(\mathbf{n}) \sim 1/r$ and, therefore, does not contribute into the long-range part of the field derivative
\be
\label{eq:TT_gauge_bh}
\left( \partial_{M} h_{ij}^{\rm TT} \right)^{\rm rad} = \Lambda_{ij,kl}(\mathbf{n}) \left( \partial_{M} \bar{h}_{kl} \right)^{\rm rad}.
\ee
However, in contrast with the four-dimensional theory, in five dimensions emitted part of the field derivative depends on the entire history of particle's motion preceding the retarded moment of proper time $\hat{\tau}$, while in four dimensions it depends only on the particle's kinematic characteristics at this moment. Nevertheless, due to the above properties of the on-shell energy-momentum tensor, the energy-momentum associated with the long-range component of the field derivative propagate at the speed of light.
 
The flux of the radiated energy-momentum passing per unit time through the  $(2\nu - 1)$-dimensional sphere of radius $r$ is given by the integral
\be
\label{eq:W_in_D_dimensions}
W_{2\nu + 1}^\mu = \int \, T_{\rm rad}^{\mu i}\; n^{i}\, r^{2\nu - 1} \, d\Omega_{2\nu - 1}, \quad i=\overline{1,2\nu},
\ee
where $d\Omega_{2\nu-1}$ is an angular element, and $\bf{n}$ is a unit spacelike vector in the direction of observation.
 
Using the Rohrlich-Teitelboim approach, the scalar synchrotron radiation from a circularly moving particle in $2 + 1$ and $4 + 1$ dimensions was calculated \cite{Galtsov:2020hhn}. The correctness of calculations was verified by the computation of the total radiation power using Fourier spectral decomposition, which are indifferent to the dimensionality of spacetime and are free from the problems discussed above.

%%%%%%%%%%%%%%%%%%%%%%%%%%%%%%%%%%%%%%%%%%%%%%%
\section{Point particles contribution}\label{III}
%%%%%%%%%%%%%%%%%%%%%%%%%%%%%%%%%%%%%%%%%%%%%%%

The point particles contribution into the gravitational waves produced by the binary system is written as
\be
\label{eq:PP_term_gen}
\bh_{MN}^{\rm P}(x) = - 2 \kappa_{5} \int d^{5}x' \, T_{MN}^{\rm P}(x') \, G_{\rm ret}^{4+1} (x - x'),
\ee
where the particles energy-momentum tensor is given by the Eq. \eqref{eq:pp_EMT_brane} and the retarded Green's function by the Eq. \eqref{eq:5D_Green}.

We are interested in the non-relativistic approximation of the emitted part of the gravitational field's \eqref{eq:PP_term_gen} derivative. To clarify the main steps of derivation, let us start with the problem of scalar radiation of non-relativistic point charge as a simple but instructive example of five-dimensional radiation of a non-relativistic source.

%%%%%%%%%%%%%%%%%%%%%%%%%%%%%%%%%%%%%%%%%%%%%%%
\subsection{Scalar radiation of a non-relativistic charge}

In this subsection, we deviate from our basic setup and consider the case of a  scalar field living in the bulk. In this case, the bulk scalar radiation (in the lowest approximation) can be calculated within the framework of a linearized theory, since there is no need to take into account the stresses that bind the system. This calculation may be interesting in itself, but for us it also provides an easier way to explain the characteristics of radiation in an odd-dimensional flat space. In fact, this gives a simplified setup for gravitational radiation, neglecting the contribution of stresses. 

By analogy with \eqref{eq:sc_EOM_flat_br}, equation of motion of a five-dimensional massless scalar field interacting with a point charge moving along an arbitrary worldline $z^{M}(\tau)$ is written as
\be
\label{eq:5D_sc_EOM}
{^{(5)}}\square \varphi(x) = 2 \pi^2 g \int d\tau \, \delta^{(5)}(x - z).
\ee
Retarded solution of Eq. \eqref{eq:5D_sc_EOM} is constructed by the corresponding five-dimensional Green's function \eqref{eq:5D_Green} and is found as
\be
\varphi(x) = g \int d\tau \, \theta(X^0) \left \lbrack \frac{1}{2} \frac{\theta(-X^2)}{(-X^2)^{3/2}} - \frac{\delta(-X^2)}{(-X^2)^{1/2}} \right \rbrack,
\ee
where we introduced the vector $X^{M}(\tau) = x^{M} - z^{M}(\tau)$ and omitted its dependence on proper time for brevity. Computing its derivative we obtain
\be
\partial_{M} \varphi(x) = g \int d\tau \, \theta(X^{0}) \left \lbrack \frac{3}{2} \frac{\theta(-X^{2})}{(-X^{2})^{5/2}} + 2 \frac{\delta^{\prime}(-X^{2})}{(-X^{2})^{1/2}} - 2 \frac{\delta(-X^{2})}{(-X^{2})^{3/2}} \right \rbrack X_{M}.
\ee
We integrate by parts the term containing derivative of delta function by use of the relation
\be
\label{eq:delta_int_parts}
\frac{dX^{2}}{d\tau} = - 2(vX),
\ee
arriving at
\begin{multline}
\label{eq:5D_scal_deriv_gen}
\partial_{M} \varphi(x) = g \int d\tau \, \theta(X^{0}) \left \lbrack \frac{3}{2} \frac{\theta(-X^{2})}{(-X^{2})^{5/2}} X_{M} - \frac{\delta(-X^{2})}{(-X^{2})^{3/2}} X_{M} + \right. \\ \left. + \frac{\delta(-X^{2})}{(vX)^{2} (-X^{2})^{1/2}} \left \lbrack (aX) + 1 \right \rbrack X_{M} - \frac{\delta(-X^{2})}{(vX) (-X^{2})^{1/2}} v_{M} \right \rbrack,
\end{multline}
where we introduced the acceleration vector $a^{M} = d^{2}z^{M}/d\tau^{2}$.

To extract the leading $\hat{\rho}$-asymptotic of the Eq. \eqref{eq:5D_scal_deriv_gen}, we use relation for the delta function of complex argument transforming its product with the Heaviside function as
\be
\label{eq:delta_compl_arg}
\theta(X^{0}) \delta(-X^{2}) = \frac{\delta(\tau - \hat{\tau})}{2\hat{\rho}}.
\ee
Also, we rewrite the vector $X^{M}$ in the following form
\be
\label{eq:X_rho_expan}
X^{M} = Z^{M} + \hat{\rho} \hat{c}^{M}, \quad Z^{M} = \hat{z}^{M} - z^{M}.
\ee
Then, expanding Eq. \eqref{eq:5D_scal_deriv_gen} in the inverse powers of the Lorentz-invariant distance $\hat{\rho}$ we find the emitted part of the scalar field derivative as
\be
\label{eq:5D_scal_deriv_emit_unreg}
\left( \partial_{M} \varphi(x) \right)^{\rm rad} = \frac{g \hat{c}_{M}}{2^{5/2} \hat{\rho}^{3/2}} \int_{-\infty}^{\hat{\tau}} d\tau \left \lbrack \frac{3}{2} \frac{1}{(-Z\hat{c})^{5/2}} - \frac{\delta(\tau - \hat{\tau})}{(-Z\hat{c})^{3/2}} + \frac{2 (a\hat{c})\delta(\tau - \hat{\tau})}{(v\hat{c})^{2} (-Z\hat{c})^{1/2}} \right \rbrack,
\ee
where we have taken into account that up to the leading order $X^{2} \sim 2 \hat{\rho} (Z\hat{c})$ and $(vX) \sim \hat{\rho} (v\hat{c})$.

Each integral in \eqref{eq:5D_scal_deriv_emit_unreg} diverges on the upper integration limit $\tau \to \hat{\tau}$, where $(Z\hat{c}) \to (\hat{v}\hat{c}) (\hat{\tau} - \tau) = - (\hat{\tau} - \tau)$. However, by use of the regularization procedure discussed in section II.D one can show that the last two terms in Eq. \eqref{eq:5D_scal_deriv_emit_unreg} do not contain physical information concerning the properties of the field in wave zone and just subtract the divergences from the first one. Indeed, we introduce the regularising parameter $\epsilon \to +0$ into the argument of delta function, obtaining the divergent result
\be
\label{eq:5D_scalar_counter-term_1}
\int_{-\infty}^{\hat{\tau}} d\tau \, \frac{\delta(\tau - \hat{\tau} + \epsilon)}{(-Z\hat{c})^{3/2}} = \frac{1}{\epsilon^{3/2}},
\ee
which is rewritten as
\be
\label{eq:5D_scalar_counter-term_2}
\frac{1}{\epsilon^{3/2}} = \frac{3}{2} \int_{-\infty}^{\hat{\tau}-\epsilon} \frac{d\tau}{(\hat{\tau} - \tau)^{3/2}}.
\ee
The second term with delta function is transformed in an analogous manner providing us with the integral
\be
\label{eq:5D_scalar_counter-term_3}
\int_{-\infty}^{\hat{\tau}} d\tau \frac{(a\hat{c}) \delta(\tau - \hat{\tau})}{(v\hat{c})^{2} (-Z\hat{c})^{1/2}} = \frac{1}{2} \int_{-\infty}^{\hat{\tau}-\epsilon} d\tau \frac{(\hat{a} \hat{c})}{(\hat{\tau} - \tau)^{3/2}}.
\ee
Then, emitted part of the retarded scalar field derivative takes the form
\be
\label{eq:5D_sc_rad}
\left( \partial_{M} \varphi(x) \right)^{\rm rad} = \lim_{\epsilon \to +0} \frac{g \hat{c}_{M}}{2^{5/2} \hat{\rho}^{3/2}} \int_{-\infty}^{\hat{\tau}-\epsilon} d\tau \left \lbrack \frac{3}{2} \frac{1}{(-Z\hat{c})^{5/2}} - \frac{3}{2} \frac{1}{(\hat{\tau} - \tau)^{5/2}} + \frac{(\hat{a}\hat{c})}{(\hat{\tau}-\tau)^{3/2}} \right \rbrack.
\ee
Further, we omit the regularising parameter in the upper integration limit for simplicity, taking in mind that to make the finiteness of the obtained expression explicit one needs to perform some transformation of the first term in the integrand: usually it is the integration by parts, which extract the divergences from the first term and subtract them with the last two integrals. The obtained structure of the emitted part of the field is the general feature of the odd-dimensional fields \cite{Galtsov:2020hhn} -- it consists of several separately divergent integrals, first of which contains all the information concerning the field in the wave zone, while the remaining ones are just the counter-terms eliminating the divergence from the first one. Note, also, that the emitted part of the retarded field depends on the entire history of the charge's motion preceding the retarded proper time $\hat{\tau}$.

Now let us find the non-relativistic approximation of the emitted part of the scalar field derivative \eqref{eq:5D_sc_rad}. We make the following assumptions about the charge's motion:
\begin{itemize}
\item
it is non-relativistic $|\mathbf{v}| \ll 1, \, \forall \tau$;
\item
it moves inside compact region of space $|\mathbf{z}| \leq d, \forall \tau$ ($d$ is the characteristic size of this region);
\item
the observation point is far from this region $d \ll r$.
\end{itemize}
Then, the covariant retarded quantities are expanded up to the leading order in small parameters $|\mathbf{v}|$ and $|\mathbf{z}|/r$ as
\begin{align}
\label{eq:htau_non-rel}
&\hat{\tau} \simeq \bar{t} + \mathbf{n}\bar{\mathbf{z}}, \quad \mathbf{n} = \mathbf{x}/r, \\
\label{eq:hrho_non-rel}
&\hat{\rho} \simeq r \left( 1 - \mathbf{n}\bar{\mathbf{v}} - \frac{1}{r} \mathbf{n}\bar{\mathbf{z}} \right), \\
\label{eq:hc_non-rel}
&\hat{c}^{M} \simeq \left \lbrack 1 + \mathbf{n}\bar{\mathbf{v}}; \, \mathbf{n} \left( 1 + \mathbf{n}\bar{\mathbf{v}} + \frac{1}{r} \mathbf{n}\bar{\mathbf{z}} \right) - \frac{1}{r}\bar{\mathbf{z}} \right \rbrack,
\end{align}
where $\bar{t} = t - r$ is the retarded time calculated up to the leading contribution, and all the barred quantities correspond to this moment. Using the Eqs. (\ref{eq:htau_non-rel}--\ref{eq:hc_non-rel}), we rewrite the first term of the integrand in Eq. \eqref{eq:5D_sc_rad} as
\be
\frac{1}{(-Z\hat{c})^{5/2}} \simeq \frac{1}{(\bar{t} - t')^{5/2}} \left \lbrack 1 + \mathbf{n}\bar{\mathbf{v}} - \mathbf{n}\mathbf{s}(t') \left( 1  + \mathbf{n}\bar{\mathbf{v}} + \frac{1}{r} \mathbf{n}\bar{\mathbf{z}} \right) - \frac{1}{r} \mathbf{s}(t')\bar{\mathbf{z}} \right \rbrack^{-5/2},
\ee
where we introduced the spacelike vector $\mathbf{s}(t')$ defined as
\be
\label{eq:s_non-rel}
\mathbf{s}(t') = \frac{\bar{\mathbf{z}} - \mathbf{z}(t')}{\bar{t} - t'},
\ee
and replaced the proper time with the coordinate one $\tau=t'$, given their equivalence in the non-relativistic limit. It is clear that at the retarded time $\bar{t}$ vector $\mathbf{s}(t')$ has a finite value $\displaystyle \lim_{t' \to \bar{t}} \mathbf{s}(t') = \bar{\mathbf{v}}$. Rewriting the particle's worldline in terms of its velocity
\be
\mathbf{z}(t) = \int_{t_{\rm in}}^{t} \mathbf{v}(t') dt' + \mathbf{z}(t_{\rm in}),
\ee
where $t_{\rm in}$ is the initial moment of time in the remote past and using the first mean value theorem, we find that the vector $\mathbf{s}(t')$ is of order of particle's velocity
\be
|\mathbf{s}(t')| \sim |\mathbf{v}|, \, \forall t'.
\ee
So, it should be used as another expansion parameter. Thus, we calculate the integrand in \eqref{eq:5D_sc_rad} up to the leading order in small parameters arriving at
\be
\label{eq:RCQ_contract_non-rel}
(-Z\hat{c})^{-5/2} = \frac{1}{(\bar{t} - t')^{5/2}} \left( 1 + \frac{5}{2} \left(\mathbf{n}\mathbf{s}(t') - \mathbf{n}\bar{\mathbf{v}} \right) \right), \quad \hat{c}_{M} = \bar{c}_{M} = \lbrack -1, \mathbf{n} \rbrack, \quad \hat{a}\hat{c} = \mathbf{n}\bar{\mathbf{a}}.
\ee
Substituting the obtained expansions into the Eq. \eqref{eq:5D_sc_rad}, we find the non-relativistic approximation of the emitted part of scalar field derivative as
\be
\label{eq:5D_sc_non-rel_gen}
(\partial_{M}\varphi(x))^{\rm rad} = - \frac{g \bar{c}_{M}}{2^{5/2} r^{3/2}} \int_{-\infty}^{\bar{t}} dt' \left \lbrack \frac{15}{4} \frac{\mathbf{n}\mathbf{s}(t')}{(\bar{t} - t')^{5/2}} - \frac{15}{4} \frac{\mathbf{n}\bar{\mathbf{v}}}{(\bar{t} - t')^{5/2}} + \frac{\mathbf{n}\bar{\mathbf{a}}}{(\bar{t} - t')^{3/2}} \right \rbrack.
\ee

To make the finitness of the emitted part of the scalar field explicit, we introduce the regularising parameter into the upper integration limit and integrate the first term in \eqref{eq:5D_sc_non-rel_gen} by parts three times arriving at
\be
\label{eq:5D_sc_lead_term_exp}
\frac{15}{4} \lim_{\epsilon \to +0} \int_{-\infty}^{\bar{t}-\epsilon} dt' \frac{\mathbf{n}\mathbf{s}(t')}{(\bar{t} - t')^{5/2}} = \frac{5}{2} \lim_{\epsilon \to +0} \frac{\mathbf{n}\bar{\mathbf{v}}}{\epsilon^{3/2}} - 2 \lim_{\epsilon \to +0} \frac{\mathbf{n}\bar{\mathbf{a}}}{\epsilon^{1/2}} + 2 \int_{-\infty}^{\bar{t}} dt' \frac{\mathbf{n}\dot{\mathbf{a}}(t')}{(\bar{t} - t')^{1/2}},
\ee
where $\dot{\mathbf{a}} = d^3 \mathbf{z} / dt^{\prime 3}$. It is clear that the first two divergent terms in Eq. \eqref{eq:5D_sc_lead_term_exp} are cancelled out by the "counter-terms" in \eqref{eq:5D_sc_non-rel_gen} yielding the simple expression for the emitted part of the scalar field derivative in the non-relativistic limit
\be
\label{eq:5D_sc_non-rel}
\left( \partial_{M} \varphi(x) \right)^{\rm rad} = - \frac{g \bar{c}_{M}}{2^{3/2} r^{3/2}} \int_{-\infty}^{\bar{t}} dt' \frac{\mathbf{n}\dot{\mathbf{a}}(t')}{(\bar{t} - t')^{1/2}}.
\ee
Note that in five-dimensions, up to the leading order in non-relativistic expansion, the rectilinear uniformly accelerated scalar charge does not radiate, in contrast with the four-dimensional theory. However, as will be shown later, this is not the case for the gravitational radiation.

Substituting the Eq. \eqref{eq:5D_sc_non-rel} into the energy-momentum tensor of the five-dimensional massless scalar field, analogous to that of Eq. \eqref{eq:sc_EMT_brane},
\be
T_{MN} = \frac{1}{2\pi^{2}} \left( \partial_{M} \varphi \partial_{N} \varphi - \frac{1}{2} \eta_{MN} \partial^{A} \varphi \partial_{A} \varphi \right),
\ee
we find the radiation energy-momentum tensor as
\be
\label{eq:5D_scal_rad_non-rel_EMT}
T_{MN}^{\rm rad} (x) = \frac{g^2 \bar{c}_{M} \bar{c}_{N}}{16 \pi^2 r^3} \left \lbrack \int_{-\infty}^{\bar{t}} dt' \frac{\mathbf{n}\dot{\mathbf{a}}(t')}{(\bar{t}-t')^{1/2}} \right \rbrack^{2},
\ee
which has the tensor structure corresponding to the propagation of the given part of energy-momentum exactly with the speed of light. One can show that in the case of charge's uniform circular motion Eq. \eqref{eq:5D_scal_rad_non-rel_EMT} yields the finite value of radiation power coinciding with the one obtained from the spectral decompositions of the radiation power indifferent to the dimensionality of spacetime found in \cite{Galtsov:2020hhn}. Note that in the non-relativistic limit the radiation power \eqref{eq:5D_scal_rad_non-rel_EMT} depends on the entire history of charge's motion preceding to the retarded time, while in the ultra-relativistic limit studied in \cite{Galtsov:2020hhn} this dependence effectively reduces to the small interval of proper time near the retarded moment.

%%%%%%%%%%%%%%%%%%%%%%%%%%%%%%%%%%%%%%%%%%%%%%%
\subsection{Emitted part of the gravitational field}

Let us return to the point particles contribution into the gravitational radiation \eqref{eq:PP_term_gen}. Due to the particles energy-momentum tensor \eqref{eq:pp_EMT_brane} being just the sum of energy-momentum tensors of individual particles, it is convenient to calculate the gravitational field of one particle and then take the sum of two analogous terms.

Retarded gravitational field of a point particle with mass $m$ moving along an arbitrary worldline $z^{M}(\tau)$, in accordance with Eq. \eqref{eq:PP_term_gen}, is given as
\be
\bh_{AB}^{\rm P} = \frac{m \kappa_{5}}{\pi^2} \int d\tau \, v_{A} v_{B} \, \theta(X^{0}) \left \lbrack \frac{\delta(-X^2)}{(-X^2)^{1/2}} - \frac{1}{2} \frac{\theta(-X^2)}{(-X^2)^{3/2}} \right \rbrack.
\ee
By analogy with the scalar field, its derivative has the form
\begin{equation}
\partial_{M} \bh_{AB}^{\rm P} = - \frac{2m \kappa_{5}}{\pi^{2}} \int d\tau \, v_{A} v_{B} \, \theta(X^{0}) \left \lbrack \frac{3}{4} \frac{\theta(-X^{2})}{(-X^{2})^{5/2}} - \frac{\delta(-X^{2})}{(-X^{2})^{3/2}} + \frac{\delta'(-X^{2})}{(-X^{2})^{1/2}} \right \rbrack X_{M}.
\end{equation}
We transform the term containing derivative of delta function by the integration by parts using the relation \eqref{eq:delta_int_parts}. We, also, use relation \eqref{eq:delta_compl_arg} to transform the products of delta and Heaviside functions. Thus, the gravitational field derivative takes the form
\begin{multline}
\label{eq:GW_deriv_gen}
\partial_{M} \bh_{AB}^{\rm P} = - \frac{m \kappa_{5}}{2 \pi^{2}} \int_{-\infty}^{\hat{\tau}} d\tau \left \lbrack 3 \frac{v_{A} v_{B} X_{M}}{(-X^2)^{5/2}} - \frac{v_{A} v_{B} X_{M}}{\hat{\rho} (-X^2)^{3/2}} \delta(\tau - \hat{\tau}) - \right. \\ - \left. \frac{2 a_{(A} v_{B)} X_{M} - v_{A} v_{B} v_{M}}{\hat{\rho} (vX) (-X^2)^{1/2}} \delta(\tau - \hat{\tau}) + \frac{v_{A} v_{B} X_{M} \lbrack (aX) + 1 \rbrack}{\hat{\rho} (vX)^{2} (-X^2)^{1/2}} \delta(\tau - \hat{\tau}) \right \rbrack,
\end{multline}
where $a^{M} = d^{2}z^{M}/d\tau^{2}$ is the acceleration five-vector, and we defined symmetrisation over two indices as $A_{(M} B_{N)} = \frac{1}{2} (A_{M} B_{N} + A_{N} B_{M})$.

We extract the emitted part of the gravitational field derivative \eqref{eq:GW_deriv_gen} by use of the relations \eqref{eq:X_rho_expan}, expanding the integrand in the inverse powers of the Lorentz-invariant distance $\hat{\rho}$. By analogy with the scalar field, it is found as
\begin{multline}
\label{eq:GW_deriv_emit_non-trans}
(\partial_{M} \bh_{AB}^{\rm P})^{\rm rad} = - \frac{m \kappa_{5} \hat{c}_{M}}{2^{3/2} \pi^{2} \hat{\rho}^{3/2}} \int_{-\infty}^{\hat{\tau}} d\tau \left \lbrack \frac{3}{4} \frac{v_{A} v_{B}}{(-Z\hat{c})^{5/2}} - \frac{1}{2} \frac{v_{A} v_{B}}{(-Z\hat{c})^{3/2}} \delta(\tau - \hat{\tau}) - \right. \\ - \left. \frac{2 a_{(A} v_{B)}}{(v\hat{c}) (-Z\hat{c})^{1/2}} \delta(\tau - \hat{\tau}) + \frac{v_{A} v_{B} (a\hat{c})}{(v\hat{c})^{2} (-Z\hat{c})^{1/2}} \delta(\tau - \hat{\tau}) \right \rbrack.
\end{multline}
Terms containing delta functions are transformed by analogy with Eqs. (\ref{eq:5D_scalar_counter-term_1} -- \ref{eq:5D_scalar_counter-term_3})
\begin{align}
&\int_{-\infty}^{\hat{\tau}} d\tau \frac{1}{2} \frac{v_{A} v_{B}}{(-Z\hat{c})^{3/2}} \delta(\tau - \hat{\tau}) = \int_{-\infty}^{\hat{\tau}} d\tau \frac{3}{4} \frac{\hat{v}_{A} \hat{v}_{B}}{(\hat{\tau} - \tau)^{5/2}}, \\
&\int_{-\infty}^{\hat{\tau}} d\tau \frac{2 a_{(A} v_{B)}}{(v\hat{c}) (-Z\hat{c})^{1/2}} \delta(\tau - \hat{\tau}) = - \int_{-\infty}^{\hat{\tau}} d\tau \, \frac{\hat{a}_{(A} \hat{v}_{B)}}{(\hat{\tau} - \tau)^{3/2}}, \\
&\int_{-\infty}^{\hat{\tau}} d\tau \frac{v_{A} v_{B} (a\hat{c})}{(v\hat{c})^{2} (-Z\hat{c})^{1/2}} \delta(\tau - \hat{\tau}) = \int_{-\infty}^{\hat{\tau}} d\tau \frac{1}{2} \frac{\hat{v}_{A} \hat{v}_{B} (\hat{a}\hat{c})}{(\hat{\tau} - \tau)^{3/2}}.
\end{align}
Therefore, the emitted part of the retarded gravitational field takes the form analogous to that of the scalar field \eqref{eq:5D_sc_rad}
\be
\label{eq:5D_GW_rad}
(\partial_{M} \bh_{AB}^{\rm P})^{\rm rad} = - \frac{m \kappa_{5} \hat{c}_{M}}{2^{5/2} \pi^{2} \hat{\rho}^{3/2}} \int_{-\infty}^{\hat{\tau}} d\tau \left \lbrack \frac{3}{2} \frac{v_{A} v_{B}}{(-Z\hat{c})^{5/2}} - \frac{3}{2} \frac{\hat{v}_{A} \hat{v}_{B}}{(\hat{\tau} - \tau)^{5/2}} + \frac{2 \hat{a}_{(A} \hat{v}_{B)} + \hat{v}_{A} \hat{v}_{B} (\hat{a}\hat{c})}{(\hat{\tau} - \tau)^{3/2}} \right \rbrack.
\ee
Note that, by analogy with the scalar field \eqref{eq:5D_sc_rad}, all the information concerning the gravitational field in the wave zone is contained in the first term of Eq. \eqref{eq:5D_GW_rad}, while the remaining terms just subtract the divergences present in the first one at the upper integration limit. Also, as in the case of scalar field, the emitted part of five-dimensional gravitational field depends on the history of the particle's motion.

%%%%%%%%%%%%%%%%%%%%%%%%%%%%%%%%%%%%%%%%%%%%%%%
\subsection{Gravitational radiation of non-relativistic particle}

Now we turn to the computation of gravitational radiation from the non-relativistic particle. By analogy with the case of scalar field, we assume that: (i) particle is non-relativistic $|\mathbf{v}| \ll 1, \, \forall \tau$; (ii) it moves inside the compact region of space $|\mathbf{z}| \leq d, \forall \tau$; (iii) the observation point is far from this region $d \ll r$. We are interested only in the spatial components of metric perturbations
\be
(\partial_{M} \bh_{ij}^{\rm P})^{\rm rad} = - \frac{m \kappa_{5} \hat{c}_{M}}{2^{5/2} \pi^{2} \hat{\rho}^{3/2}} \int_{-\infty}^{\hat{\tau}} d\tau \left \lbrack \frac{3}{2} \frac{v_{i} v_{j}}{(-Z\hat{c})^{5/2}} - \frac{3}{2} \frac{\hat{v}_{i} \hat{v}_{j}}{(\hat{\tau} - \tau)^{5/2}} + \frac{2 \hat{a}_{(i} \hat{v}_{j)} + \hat{v}_{i} \hat{v}_{j} (\hat{a}\hat{c})}{(\hat{\tau} - \tau)^{3/2}} \right \rbrack,
\ee
because it is them, which contribute into the effective energy-momentum tensor of gravitational field \eqref{eq:GW_EMT_gen}.

By analogy with the scalar field, using expansions of the retarded covariant quantities Eqs. (\ref{eq:htau_non-rel}--\ref{eq:hc_non-rel}) and introducing the spacelike vector $\mathbf{s}(t')$ defined by Eq. \eqref{eq:s_non-rel} (here we replaced the proper time with coordinate one), we obtain the non-relativistic approximation of the emitted part of the gravitatonal field up to the leading order in small parameters $|\mathbf{z}|/r, |\mathbf{v}|, |\mathbf{s}| \ll 1$ as
\begin{multline}
\label{eq:pp_GW_non-rel_non-int}
(\partial_{M}\bh_{ij}^{\rm P})^{\rm rad} = - \frac{m \kappa_{5} \bar{c}_{M}}{2^{5/2} \pi^2 r^{3/2}} \int_{-\infty}^{\bar{t}} dt' \left \lbrack \frac{15}{4} v_{i} v_{j} \frac{\mathbf{n}\mathbf{s} - \mathbf{n}\bar{\mathbf{v}}}{(\bar{t} - t')^{5/2}} + \frac{3}{2} \frac{v_{i} v_{j}}{(\bar{t} - t')^{5/2}} - \frac{3}{2} \frac{\bar{v}_{i} \bar{v}_{j}}{(\bar{t} - t')^{5/2}} + \right. \\ + \left. \frac{2 \bar{a}_{(i} \bar{v}_{j)} + \bar{v}_{i} \bar{v}_{j} \mathbf{n}\bar{\mathbf{a}}}{(\bar{t} - t')^{3/2}} \right \rbrack.
\end{multline}
Here, all the information concerning the gravitational field in the wave zone is contained in the first two terms of the integrand, while the remaining two terms just subtract the divergences present in them at the upper integration limit.

To make the convergence of integral \eqref{eq:pp_GW_non-rel_non-int} explicit, by analogy with the scalar field, one needs to integrate the first two terms of the integrand by parts reducing the powers of denominators to $1/2$. Detailed discussion of this integration is relegated to the Appendix B, while here we just present the result remaining after the mutual cancellation of divergences
\begin{multline}
\label{eq:pp_GW_non-rel_mixed}
(\partial_{M}\bh_{ij}^{\rm P})^{\rm rad} = - \frac{m \kappa_{5} \bar{c}_{M}}{2^{5/2} \pi^2 r^{3/2}} \int_{-\infty}^{\bar{t}} dt' \left \lbrack 2 v_{i} v_{j} \frac{\mathbf{n}\dot{\mathbf{a}}}{(\bar{t} - t')^{1/2}} + 12 a_{(i} v_{j)} \frac{\mathbf{n}\mathbf{a}}{(\bar{t} - t')^{1/2}} + \right. \\ + \left. \left( \dot{a}_{i} v_{j} + 2 a_{i} a_{j} + v_{i} \dot{a}_{j} \right) \frac{6\mathbf{n}\mathbf{v} - 5\mathbf{n}\bar{\mathbf{v}} + 2}{(\bar{t} - t')^{1/2}} - 2 \left( \ddot{a}_{i} v_{j} + 3 \dot{a}_{i} a_{j} + 3 a_{i} \dot{a}_{j} + v_{i} \ddot{a}_{j} \right) \frac{\mathbf{n}\bar{\mathbf{z}} - \mathbf{n}\mathbf{z}}{(\bar{t} - t')^{1/2}} \right \rbrack.
\end{multline}
However, the terms in the integrand of Eq. \eqref{eq:pp_GW_non-rel_mixed} does not contribute equivalently into the non-relativistic approximation of the emitted part of the gravitational field.

To perform power counting, let us consider the Fourier transform of the particle's world line
\be
z^{k}(t) = \int \frac{d\omega}{2\pi} \, \tilde{z}^{k}(\omega) e^{-i\omega t}, \quad \tilde{z}^{k}(\omega) = \int dt \, z^{k}(t) e^{i\omega t}.
\ee
When particle performs a periodic motion inside the compact region of space $|\mathbf{z}| \leq d$, the Fourier transform of its world line satisfies the inequality
\be
\tilde{z}^{k}(\omega) \leq d \int dt e^{i\omega t} = 2\pi d \delta(\omega),
\ee
where the right-hand side corresponds to the particle at rest at the border of this region. Thus, the Fourier transform of periodically moving particle's world line has to be of form
\be
\tilde{z}^{k}(\omega) = 2\pi d f^{k}(\omega),
\ee
where $f^{k}(\omega)$ are the functions localised around a characteristic frequency of periodic motion $\omega_{\rm ch}$. Then, the $n$-th derivative of the particle's world line is found as
\be
\frac{d^{n} z^{k}}{dt^{n}} = (-i)^{n} \int d\omega \, d \, \omega^{n} f^{k}(\omega) e^{-i\omega t} \sim - i \frac{|\mathbf{v}|}{d} \frac{d^{n-1}z^{k}}{dt^{n-1}},
\ee
where, due to the localisation of integration interval around $\omega_{\rm ch}$, product $d \omega$ has to be of order of particle's velocity $|\mathbf{v}|$. The case of circular motion is the most transparent example
\begin{align}
&z^{k}(t) = R_{0} \cos(\omega_{0}t + \phi), \\
&\frac{d^{n} z^{k}}{dt^{n}} = \omega^{n-1}_{0} |\mathbf{v}| \cos\left(\omega_{0}t + \phi + \frac{\pi n}{2}\right),
\end{align}
where $\omega_{0} R_{0}$ is the particle's velocity. Therefore, in the non-relativistic limit, all the particle's kinematic characteristics are of order of its velocity $d^{n}z^{k}/dt^{n} \sim |\mathbf{v}| \ll 1$. Following the same reasoning, one can show that $z^{k}\, d^{n}z^{l}/dt^{n} \sim |\mathbf{v}|^{2}, \, n \geq 2$.

Based on the power counting scheme above, we find that in the non-relativistic limit only one term in the integrand in Eq. \eqref{eq:pp_GW_non-rel_mixed} contributes into the emitted part of the gravitational field
\be
\label{eq:pp_GW_rad_non-rel}
(\partial_{M}\bh_{ij}^{\rm P})^{\rm rad} = - \frac{m \kappa_{5} \bar{c}_{M}}{2^{3/2} \pi^2 r^{3/2}} \int_{-\infty}^{\bar{t}} dt' \frac{\dot{a}_{i} v_{j} + 2 a_{i} a_{j} + v_{i} \dot{a}_{j}}{(\bar{t} - t')^{1/2}}.
\ee
Note that, by analogy with the scalar field \eqref{eq:5D_sc_non-rel}, in the non-relativistic limit, emitted part of the gravitational field depends on the entire history of particle's motion preceding the retarded time.

%%%%%%%%%%%%%%%%%%%%%%%%%%%%%%%%%%%%%%%%%%%%%%%
\subsection{Gravitational radiation of binary system}

Having found the non-relativistic approximation of the gravitational field of individual particle \eqref{eq:pp_GW_rad_non-rel}, we proceed to the case of non-relativistic binary system. Without loss of generality, all the previous calculations could be performed in the center-of-mass frame
\be
\label{eq:COM_frame_def}
\mathbf{x}_{\rm CM} = \frac{m_{1}\mathbf{z}_{1} + m_{2}\mathbf{z}_{2}}{M} = 0, \forall t,
\ee
where $M=m_1 + m_2$ is the total mass of the system. Then, the particles' coordinates are rewritten in terms of the relative coordinate $\mathbf{z}=\mathbf{z}_2 - \mathbf{z}_1$ in a standard manner
\be
\label{eq:rel_coord_def}
\mathbf{z}_{1} = - \frac{m_2}{M} \mathbf{z}, \quad \mathbf{z}_{2} = \frac{m_1}{M} \mathbf{z}.
\ee
Therefore, in the center-of-mass frame, superposition of gravitational field's from individual particles forming the binary system is given by the integral \eqref{eq:pp_GW_rad_non-rel} calculated on the relative coordinate
\be
\label{eq:pp_GW_rad_non-rel_COM}
(\partial_{M} \bh_{ij}^{\rm P})^{\rm rad} = - \frac{\mu \kappa_{5} \bar{c}_{M}}{2^{3/2} \pi^2 r^{3/2}} \int_{-\infty}^{\bar{t}} dt' \frac{\dot{a}_{i} v_{j} + 2 a_{i} a_{j} + v_{i} \dot{a}_{j}}{(\bar{t} - t')^{1/2}},
\ee
where $\mu = m_{1}m_{2}/M$ is the reduced mass of the system. It is the Eq. \eqref{eq:pp_GW_rad_non-rel_COM} that gives the point particles contribution into the gravitational radiation of the binary system.

%%%%%%%%%%%%%%%%%%%%%%%%%%%%%%%%%%%%%%%%%%%%%%%
\section{Scalar field contribution}\label{IV}
%%%%%%%%%%%%%%%%%%%%%%%%%%%%%%%%%%%%%%%%%%%%%%%

Contribution of the scalar field into the gravitational radiation of the binary system, by analogy with the one from the point particles \eqref{eq:PP_term_gen}, is written as
\be
\label{eq:SF_term_gen}
\bh_{MN}^{\rm F}(x) = - 2 \kappa_{5} \int d^{5}x' \, T_{MN}^{\rm F}(x') \, G_{\rm ret}^{4+1} (x - x'),
\ee
where the scalar field energy-momentum tensor is given by Eq. \eqref{eq:sc_EMT_brane} and the retarded Green's function is defined by Eq. \eqref{eq:5D_Green}.

Recall that to determine the energy-momentum flux of gravitational radiation \eqref{eq:GW_EMT_gen} one needs only the spatial components of metric perturbations in the transverse-traceless gauge. Computing the contractions \eqref{eq:TT_gauge_practical}, we take into account the vanishing traces of the projector \eqref{eq:Lambda_traces} and that the second term in the scalar field energy-momentum tensor $T_{ij}^{\rm F}$ is proportional to the Kronecker delta. Therefore, the scalar field energy-momentum tensor effectively reduces to
\be
\label{eq:sc_EMT_brane_spatial}
T_{ij}^{\rm F} = \frac{1}{4\pi} \partial_{i} \varphi \partial_{j} \varphi \, \delta(x^{4}).
\ee
Also, recall the condition for absence of the scalar radiation from the system. Based on it, we omit the self-energy type terms in the scalar field energy-momentum tensor
\be
T_{ij}^{\rm F} = \frac{1}{4\pi} \partial_{i} \varphi_{1} \partial_{j} \varphi_{2} \, \delta(x^{4}) + (i \leftrightarrow j)
\ee
leaving only the interaction type ones. Therefore, we obtain the effective scalar field contribution into the gravitational field as
\begin{align}
\label{eq:sc_term_effective}
&\bh_{ij}^{\rm F}(x) = J_{ij}(x) + (i \leftrightarrow j), \\
\label{eq:J_int_def}
&J_{ij}(x) = - \frac{\kappa_{5}}{2\pi} \int d^{5}x' \, \partial_{i}^{\,\prime} \varphi_{1}(x') \, \partial_{j}^{\,\prime} \varphi_{2}(x') \, G_{\rm ret}^{5} (x - x') \, \delta(x^{\prime 4}).
\end{align}
We compute the $J_{ij}(x)$ integral and symmetrise the obtained result over the indices to find the gravitational radiation produced by the scalar field.

\begin{figure}[t]
\begin{minipage}[h]{0.49\linewidth}
\center{\includegraphics[width=0.6\linewidth]{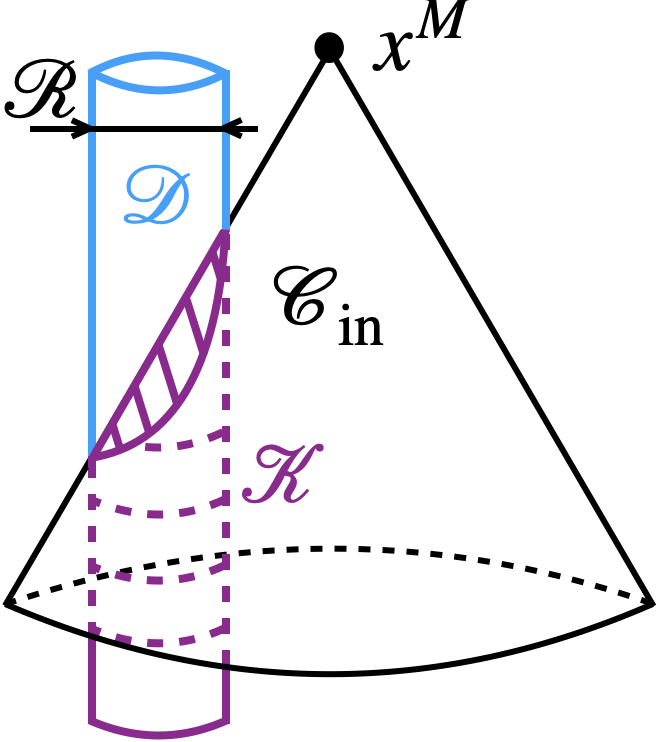}}
\end{minipage}
\hfill
\begin{minipage}[h]{0.49\linewidth}
\center{\includegraphics[width=0.6\linewidth]{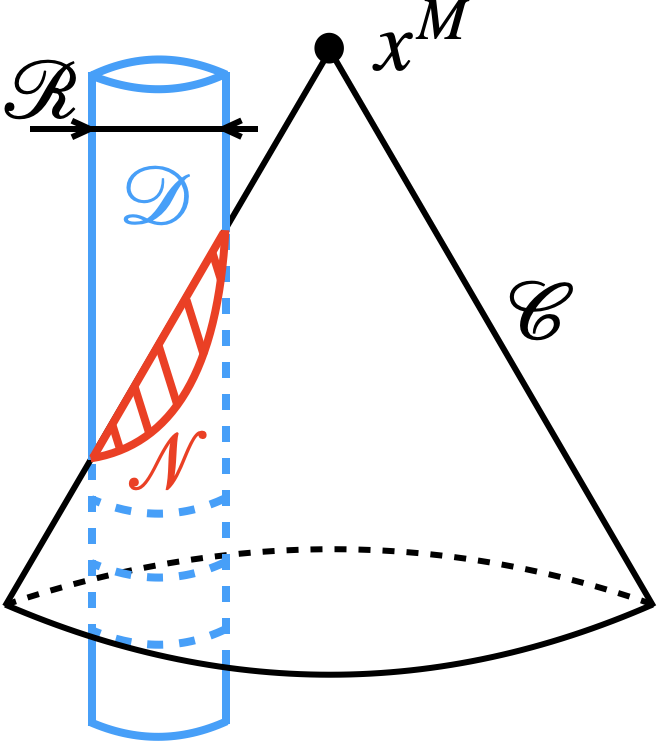}}
\end{minipage}
\caption{The spacetime splitting into the near and radiation zones. Integration region for the tail integral ${\cal C}_{\rm in}$ and its intersection with the near zone $\cal K$ (left figure). Integration region for the cone integral ${\cal C}$ and its intersection with the near zone $\cal N$ (right figure).}
\label{fig:1}
\end{figure}

Given the structure of retarded Green's function \eqref{eq:5D_Green}, we split the $J_{ij}(x)$ integral into two terms with different geometrical interpretations
\begin{align}
&J_{ij}(x) = J_{ij}^{\rm T}(x) + J_{ij}^{\rm C}(x), \\
\label{eq:tail_int_def}
&J_{ij}^{\rm T}(x) = - \frac{\kappa_{5}}{8 \pi^{3}} \int d^{5}x' \, T_{ij}(x^{\prime}) \, \theta(X^0) \, \theta(-X^2) \, \delta(x^{\prime 4}), \\
\label{eq:cone_int_def}
&J_{ij}^{\rm C}(x) = \frac{\kappa_{5}}{4 \pi^{3}} \int d^{5}x' \, C_{ij}(x^{\prime}) \, \theta(X^0) \delta(-X^2) \, \delta(x^{\prime 4}),
\end{align}
where we introduced vector $X^M=x^M-x^{\prime M}$. The tail integral \eqref{eq:tail_int_def} corresponds to the integration of tensor
\be
T_{ij}(x') = \frac{\partial_{i}^{\,\prime} \varphi_{1}(x') \, \partial_{j}^{\,\prime} \varphi_{2}(x')}{(-X^2)^{3/2}}
\ee
over the cross-section of the light cone in the past with top at the observation point $x^{M}$ and of its interior by the hypersurface $x^{\prime 4}=0$. The corresponding integration region is denoted as ${\cal C}_{\rm in}$, see Fig. \eqref{fig:1}. Analogously, the cone integral \eqref{eq:cone_int_def} corresponds to the integration of tensor
\be
\label{eq:C_ten_def}
C_{ij}(x^{\prime}) = \frac{\partial_{i}^{\,\prime} \varphi_{1}(x') \, \partial_{j}^{\,\prime} \varphi_{2}(x')}{(-X^{2})^{1/2}}
\ee
over the cross-section of the light cone by the hypersurface $x^{\prime 4} = 0$. This integration region is denoted as $\cal C$, see Fig. \eqref{fig:1}. Further, we demonstrate that, by analogy with the point particles contribution, the tail integral contains all the information about the gravitational field in the wave zone, while the cone one is just the counter-term subtracting the divergences contained in the former on the light cone.

%%%%%%%%%%%%%%%%%%%%%%%%%%%%%%%%%%%%%%%%%%%%%%%
\subsection{Spacetime splitting -- near and radiation zones}

In our computations we follow the DIRE approach to the post-Newtonian expansions in the four-dimensional GR \cite{Pati:2000vt} (see, also, \cite{Thorne:1980ru,Will:1996zj} for compendium of formulas), limiting ourselves to the leading order contributions. In this approach, splitting the spacetime into zones yielding different in orders of magnitude contributions into the gravitational field seems to be essential. So, let us briefly recall their definitions.

In what follows, we work in the center-of-mass frame \eqref{eq:COM_frame_def} and denote the characteristic size of the binary system as ${\cal S}$. First, the {\it source zone} ${\cal T}$ is defined as the world tube covering the region of particles motion
\be
\label{eq:source_zone_def}
{\cal T} = \left \lbrace x^{M} \vert r < {\cal S}; -\infty < x^{0} < +\infty \right \rbrace.
\ee
It is clear that outside the source zone $\cal T$ particles energy-momentum tensor vanishes $T_{MN}^{\rm P} = 0$. In the center-of-mass frame, the origin lies inside $\cal T$.

We assume that the non-relativistic particles move with characteristic velocity $v \ll 1$ determining the characteristic wavelength $\lambdabar_{\rm GW} \sim {\cal S}/v \gg {\cal S}$ of the gravitational waves generated by the system. Then, the {\it near zone} $\cal D$ is defined as the world tube of radius ${\cal R} = {\cal S}/v$
\be
\label{eq:near_zone_def}
{\cal D} = \left \lbrace x^{M} \vert r < {\cal R}; -\infty < x^{0} < +\infty \right \rbrace,
\ee
see Figs. \eqref{fig:1}. Inside $\cal D$, we consider the gravitational and scalar fields as the instantaneous functions of the particles world lines, i.e., we neglect the retardation of the fields inside the near zone. Finally, the {\it radiation zone} is defined as the region exterior to the near zone, $r>{\cal R}$. As we are interested in the gravitational radiation of the system, the observation point is assumed to lie in the radiation zone, $|\mathbf{x}| \gg {\cal R}$.

Based on the spacetime zones above, we split the integration regions of the tail and cone integrals. We denote the intersection of the near zone with the tail integral's region as ${\cal K} = {\cal C}_{\rm in} \cap {\cal D}$, see Fig. \eqref{fig:1}. Analogously, intersection of the near zone with the cone integral's hypersurface is denoted as ${\cal N} = {\cal C} \cap {\cal D}$, see Fig. \eqref{fig:1}. Then, the Eqs. \eqref{eq:tail_int_def} and \eqref{eq:cone_int_def} split into the integrals over the near and radiation zones
\begin{align}
&J_{ij}^{\rm T}(x) = - \frac{\kappa_{5}}{8 \pi^{3}} \int_{{\cal K}} d^{5}x' \, T_{ij}(x^{\prime}) - \frac{\kappa_{5}}{8 \pi^{3}} \int_{{\cal C}_{\rm in}\setminus{\cal K}} d^{5}x' \, T_{ij}(x^{\prime}), \\
&J_{ij}^{\rm C}(x) = \frac{\kappa_{5}}{4 \pi^{3}} \int_{{\cal N}} d^{5}x' \, C_{ij}(x^{\prime}) + \frac{\kappa_{5}}{4 \pi^{3}} \int_{{\cal C}\setminus{\cal N}} d^{5}x' \, C_{ij}(x^{\prime}).
\end{align}
However, as we assume particles to be non-relativistic and we are interested only in the leading contributions into the tail and cone integrals, we suppose the radius of the near zone $\cal R$ to be large enough for the scalar field energy-momentum tensor $T_{ij}^{\rm F}$ to be vanishingly small in the radiation zone. Thus, the tail and cone integrals effectively reduce to
\be
\label{eq:tail_cone_ints_reduced}
J_{ij}^{\rm T}(x) = - \frac{\kappa_{5}}{8 \pi^{3}} \int_{{\cal K}} d^{5}x' \, T_{ij}(x^{\prime}), \quad J_{ij}^{\rm C}(x) = \frac{\kappa_{5}}{4 \pi^{3}} \int_{{\cal N}} d^{5}x' \, C_{ij}(x^{\prime}).
\ee
Also, based on the reasoning above, we suppose that the boundary terms, which would arise from the integrations by parts of the spatial integrals in \eqref{eq:tail_cone_ints_reduced}, are defined by regions $|\mathbf{x}'| \sim {\cal R}$ and, thus, negligibly small. By analogy, we neglect the contributions into resulting integrals proportional to ${\cal R}^{-n}, n>0$.

%%%%%%%%%%%%%%%%%%%%%%%%%%%%%%%%%%%%%%%%%%%%%%%
\subsection{Tail integral}

We start with computation of the tail integral \eqref{eq:tail_cone_ints_reduced} and its contribution into the emitted part of gravitational field. Restoring its explicit form we arrive at
\be
\label{eq:tail_int_pract_start}
J_{ij}^{\rm T}(x) = - \frac{\kappa_{5}}{8 \pi^{3}} \int_{{\cal K}} d^{4}x' \, \partial_{i}^{\,\prime} \varphi_{1}(x') \, \partial_{j}^{\,\prime} \varphi_{2}(x') \, \theta(X^0) \left. \frac{\theta(-X^2)}{(-X^2)^{3/2}} \right \vert_{x^{\prime  4} = 0},
\ee
where the subscript $\cal K$ denotes that we limit integration region to the near zone, and we performed integration over the $x^{\prime 4}$-coordinate by use of the delta function.

Heaviside functions in Eq. \eqref{eq:tail_int_pract_start} determine the interval for the temporal integral
\be
\label{eq:ret_time}
t' \leq t - \sqrt{(\vec{x} - \vec{x}')^{2} + (x^{4})^{2}} = t_{\rm ret}(\vec{x}'),
\ee
where $t=x^0$, and we denote coordinates on the brane as $\vec{x}$. Separating the spatial and temporal integrals, we obtain the tail integral as
\be
\label{eq:tail_used_Heaviside}
J_{ij}^{\rm T}(x) = - \frac{\kappa_{5}}{8 \pi^{3}} \int_{{\cal K}} d^{3}x^{\prime} \int_{-\infty}^{t_{\rm ret}} dt' \frac{\partial_{i}^{\,\prime} \varphi_{1}(x') \, \partial_{j}^{\,\prime} \varphi_{2}(x')}{\left \lbrack (t - t')^{2} - (\vec{x} - \vec{x}')^{2} - (x^{4})^{2} \right \rbrack^{3/2}}.
\ee
Note that for the observation point in the radiation zone $|\vec{x}'| \leq {\cal R} \ll r$. Then, expanding the retarded time \eqref{eq:ret_time} and the denominator in Eq. \eqref{eq:tail_used_Heaviside} up to the leading order in small parameter $|\vec{x}'|/r \ll 1$ as
\begin{align}
&t_{\rm ret}(\vec{x}') \simeq \bar{t} + \vec{x} \vec{x}'/r, \\
&(t - t')^{2} - (\vec{x} - \vec{x}')^{2} - (x^{4})^{2} \simeq (t - t')^{2} - r^{2} + 2 \vec{x} \vec{x}',
\end{align}
we come to the tail integral in form
\be
\label{eq:tail_non_expand}
J_{ij}^{\rm T}(x) = - \frac{\kappa_{5}}{8 \pi^{3}} \int_{{\cal K}} d^{3}x^{\prime} \int_{-\infty}^{\bar{t} + \vec{x} \vec{x}'/r} dt' \frac{\partial_{i}^{\,\prime} \varphi_{1}(x') \, \partial_{j}^{\,\prime} \varphi_{2}(x')}{\left \lbrack (t - t')^{2} - r^{2} + 2 \vec{x} \vec{x}' \right \rbrack^{3/2}}.
\ee
As we are interested only in the leading contribution into the tail and cone integrals, we expand the temporal integral in \eqref{eq:tail_non_expand} up to the leading order in $|\vec{x}'|/r$, obtaining
\be
\label{eq:tail_int_lead}
J_{ij}^{\rm T}(x) = - \frac{\kappa_{5}}{8 \pi^{3}} \int_{{\cal K}} d^{3}x^{\prime} \left \lbrack \int_{-\infty}^{\bar{t}} dt' \frac{\partial_{i}^{\,\prime} \varphi_{1}(x') \, \partial_{j}^{\,\prime} \varphi_{2}(x')}{\lbrack (t - t')^{2} - r^{2} \rbrack^{3/2}} + {\cal O}\left( \frac{\vec{x}'}{r} \right) \right \rbrack.
\ee
Therefore, in Eq. \eqref{eq:tail_int_lead} the spatial and temporal integrals decoupled and we can calculate them independently.

To transform the Eq. \eqref{eq:tail_int_lead} into the form analogous to that of the point particles contribution \eqref{eq:pp_GW_rad_non-rel_COM}, we have to compute the spatial integral. We integrate it by parts
\be
\label{eq:tail_int_separated}
J_{ij}^{\rm T}(x) = \frac{\kappa_{5}}{8 \pi^{3}} \int_{{\cal K}} d^{3}x^{\prime} \int_{-\infty}^{\bar{t}} dt' \frac{\varphi_{1}(x') \partial_{i}^{\,\prime} \partial_{j}^{\,\prime} \varphi_{2}(x')}{\lbrack (t - t')^{2} - r^{2} \rbrack^{3/2}},
\ee
omitting, as discussed above, the boundary terms. Given that inside $\cal K$ retardation of the fields is negligible, the scalar fields are given by Eq. \eqref{eq:ret_sc_non-rel}, and the obtained integral is easily computed using the methods developed in the DIRE approach. We relegate the detailed discussion of this calculation to the Appendix C, while here we just present the result
\be
\label{eq:spatial_int}
\int_{{\cal K}} d^{3}x^{\prime} \, \varphi_{1}(x') \partial_{i}^{\,\prime} \partial_{j}^{\,\prime} \varphi_{2}(x') = - \pi \mu \lbrack a_{i}(t') z_{j}(t') +  z_{i}(t') a_{j}(t') \rbrack
\ee
in terms of the relative coordinate of the binary system $z_{i}(t')$, where $\mu$ is its reduced mass. Thus, the tail integral takes the form
\be
\label{eq:tail_int_no_spatial}
J_{ij}^{\rm T}(x) = - \frac{\mu \kappa_{5}}{8 \pi^{2}} \int_{-\infty}^{\bar{t}} dt' \frac{a_{i} z_{j} + z_{i} a_{j}}{\lbrack (t - t')^{2} - r^{2} \rbrack^{3/2}}.
\ee
Also, we transform the denominator in the Eq. \eqref{eq:tail_int_no_spatial}, rewriting the observation time $t$ in terms of the retarded time $t = \bar{t} + r$ and arriving at
\be
\label{eq:tail_int_denom_trans}
J_{ij}^{\rm T}(x) = - \frac{\mu \kappa_{5}}{8 \pi^{2}} \int_{-\infty}^{\bar{t}} dt' \frac{a_{i} z_{j} + z_{i} a_{j}}{(\bar{t} - t')^{3/2} \lbrack \bar{t} - t' + 2r \rbrack^{3/2}}.
\ee
It is clear that the leading contribution into this integral comes from the small region around the retarded time $\bar{t} - t' \ll 2r$. Therefore, the leading contribution into the tail integral has the form
\be
\label{eq:tail_int_result}
J_{ij}^{\rm T}(x) = - \frac{\mu \kappa_{5}}{2^{9/2} \pi^{2} r^{3/2}} \int_{-\infty}^{\bar{t}} dt' \, \frac{a_{i} z_{j} + z_{i} a_{j}}{(\bar{t} - t')^{3/2}}
\ee
analogous to the contribution of the point particles into the gravitational field of the system \eqref{eq:pp_GW_rad_non-rel_COM}. Note that the obtained integral diverges on the upper integration limit.

%%%%%%%%%%%%%%%%%%%%%%%%%%%%%%%%%%%%%%%%%%%%%%%
\subsection{Cone integral}

Now we turn to the calculation of the cone integral \eqref{eq:tail_cone_ints_reduced}, which explicitly is written as
\be
J_{ij}^{\rm C}(x) = \frac{\kappa_{5}}{4 \pi^{3}} \int_{{\cal N}} d^{4}x' \, \partial_{i}^{\,\prime} \varphi_{1}(x') \, \partial_{j}^{\,\prime} \varphi_{2}(x') \, \theta(X^0) \left. \frac{\delta(-X^2)}{(-X^2)^{1/2}} \right \vert_{x^{\prime 4}=0},
\ee
where we performed the integration over $x^{\prime 4}$-coordinate by use of the delta function.

We transform the product of delta and Heaviside functions by use of the formula for delta function of complex argument
\be
\left. \theta(X^0) \delta(-X^2) \right \vert_{x^{\prime 4}=0} = \frac{\delta(t' - t_{\rm ret})}{2\sqrt{(\vec{x} - \vec{x}')^2 + (x^{4})^{2}}},
\ee
where $t_{\rm ret}$ is the retarded time defined by Eq. \eqref{eq:ret_time}. Thus, the cone integral takes the form
\be
\label{eq:cone_int_simp_delt}
J_{ij}^{\rm C}(x) = \frac{\kappa_{5}}{8 \pi^{3}} \int_{{\cal N}} d^{3}x' \int dt' \, \frac{\partial_{i}^{\,\prime} \varphi_{1}(x') \, \partial_{j}^{\,\prime} \varphi_{2}(x') \, \delta(t' - t_{\rm ret})}{\sqrt{(\vec{x} - \vec{x}')^{2} + (x^{4})^{2}} \sqrt{(t - t')^{2} - (\vec{x} - \vec{x}')^{2} - (x^{4})^{2}}}.
\ee
By analogy with the tail integral, we expand the Eq. \eqref{eq:cone_int_simp_delt} with respect to the small parameter $|\vec{x}'|/r \ll 1$ keeping only the leading contribution
\be
\label{eq:cone_int_lead_multipole}
J_{ij}^{\rm C}(x) = \frac{\kappa_{5}}{8 \pi^{3} r} \int_{{\cal N}} d^{3}x' \int dt' \, \frac{\partial_{i}^{\,\prime} \varphi_{1}(x') \, \partial_{j}^{\,\prime} \varphi_{2}(x') \, \delta(t' - \bar{t})}{\sqrt{(t - t')^{2} - r^{2}}},
\ee
where the spatial and temporal integrals decouple.

Spatial integral in \eqref{eq:cone_int_lead_multipole} is equivalent to the the one obtained in the computation of the tail integral and is given by Eq. \eqref{eq:spatial_int}. Thus, the cone is written as
\be
\label{eq:cone_int_non-simp}
J_{ij}^{\rm C}(x) = \frac{\mu \kappa_{5}}{8 \pi^{2} r} \int dt' \, \frac{a_{i} z_{j} + z_{i} a_{j}}{\sqrt{(t - t')^{2} - r^{2}}} \, \delta(t' - \bar{t}),
\ee
where $z^{i}(t')$ is the relative coordinate of the binary system. Rewriting the coordinate time in terms of the retarded time $t = \bar{t} + r$ and taking into account that the leading contribution into the integral \eqref{eq:cone_int_non-simp} comes from the point $t' = \bar{t}$, due to the delta function, we transform denominator of the integrand by analogy with Eq. \eqref{eq:tail_int_result}, obtaining
\be
\label{eq:cone_int_with_delta}
J_{ij}^{\rm C}(x) = \frac{\mu \kappa_{5}}{2^{7/2} \pi^{2} r^{3/2}} \int dt' \, \frac{a_{i} z_{j} + z_{i} a_{j}}{(\bar{t} - t')^{1/2}} \, \delta(t' - \bar{t}).
\ee
Transforming the Eq. \eqref{eq:cone_int_with_delta} by analogy with Eqs. \eqref{eq:5D_scalar_counter-term_1} and \eqref{eq:5D_scalar_counter-term_2}, we come the cone integral in the form
\be
\label{eq:cone_int_result}
J_{ij}^{\rm C}(x) = \frac{\mu \kappa_{5}}{2^{9/2} \pi^{2} r^{3/2}} \int_{-\infty}^{\bar{t}} dt' \, \frac{\bar{a}_{i} \bar{z}_{j} + \bar{z}_{i} \bar{a}_{j}}{(\bar{t} - t')^{3/2}}.
\ee
It is clear that, by analogy with the point patricles contribution, the cone integral \eqref{eq:cone_int_with_delta} is just the counter-term subtracting the divergence contained in the tail integral \eqref{eq:tail_int_result} at the upper integration limit.

%%%%%%%%%%%%%%%%%%%%%%%%%%%%%%%%%%%%%%%%%%%%%%%
\subsection{Emitted part of the gravitational field}

Combining Eqs. \eqref{eq:tail_int_result} and \eqref{eq:cone_int_result} and taking into account Eq. \eqref{eq:sc_term_effective}, we find the scalar field's contribution into the gravitational radiation of the system
\be
\label{eq:scal_grav_field_gen}
\bh_{ij}^{\rm F}(x) = - \frac{\mu \kappa_{5}}{2^{7/2} \pi^{2} r^{3/2}} \int_{-\infty}^{\bar{t}} dt' \left \lbrack \frac{a_{i} z_{j} + z_{i} a_{j}}{(\bar{t} - t')^{3/2}} - \frac{\bar{a}_{i} \bar{z}_{j} + \bar{z}_{i} \bar{a}_{j}}{(\bar{t} - t')^{3/2}} \right \rbrack.
\ee
In accordance with the Rohrlich-Teitelboim approach, one needs to find the emitted part of the gravitational field derivative. Note that all the observation point dependence of the Eq. \eqref{eq:scal_grav_field_gen} is contained in the prefactor, proportional to the inverse distance $1/r$, and in the retarded time $\bar{t}$. Differentiation of the prefactor increases the decay power of  the field asymptotic and, thus, does not contribute to the emitted part. In addition, differentiation of the upper limit of integration containing retarded time leads to the sum of two diverging terms that cancel each other out, and thus also do  not contribute to gravitational radiation. 
Therefore, the emitted part of the gravitational field produced by the scalar field stresses has the form
\begin{multline}
\label{eq:scal_grav_emit_gen}
(\partial_{M} \bh_{ij}^{\rm F})^{\rm rad} = \frac{\mu \kappa_{5} (\partial_{M}\bar{t})^{\rm rad}}{2^{7/2} \pi^{2} r^{3/2}} \int_{-\infty}^{\bar{t}} dt' \left \lbrack \frac{3}{2} \frac{a_{i} z_{j} + z_{i} a_{j}}{(\bar{t} - t')^{5/2}} - \frac{3}{2} \frac{\bar{a}_{i} \bar{z}_{j} + \bar{z}_{i} \bar{a}_{j}}{(\bar{t} - t')^{5/2}} + \right. \\ + \left. \frac{\bar{\dot{a}}_{i} \bar{z}_{j} + \bar{a}_{i} \bar{v}_{j} + \bar{v}_{i} \bar{a}_{j} + \bar{z}_{i} \bar{\dot{a}}_{j}}{(\bar{t} - t')^{3/2}} \right \rbrack.
\end{multline}
By analogy with the contribution of point  particles, the first term in the Eq. \eqref{eq:scal_grav_emit_gen} contains all the physical information concerning the gravitational field in the wave zone, while two remaining terms are just the counter-terms subtracting the divergences contained in the first one at the upper integration limit. By analogy with Eq. \eqref{eq:5D_sc_lead_term_exp}, to make the convergence of the integral \eqref{eq:scal_grav_emit_gen} explicit, we introduce the regularising parameter $\epsilon \to +0$ into the upper integration limit $\bar{t} \to \bar{t} - \epsilon$ and integrate the first term by parts twice, arriving at
\begin{multline}
\label{eq:sc_res_int}
\lim_{\epsilon \to +0} \frac{3}{2} \int_{-\infty}^{\bar{t} - \epsilon} dt' \frac{a_{i} z_{j} + z_{i} a_{j}}{(\bar{t} - t')^{5/2}} = \lim_{\epsilon \to +0} \frac{\bar{a}_{i} \bar{z}_{j} + \bar{z}_{i} \bar{a}_{j}}{\epsilon^{3/2}} - 2 \lim_{\epsilon \to +0} \frac{\bar{\dot{a}}_{i} \bar{z}_{j} + \bar{a}_{i} \bar{v}_{j} + \bar{v}_{i} \bar{a}_{j} + \bar{z}_{i} \bar{\dot{a}}_{j}}{\epsilon^{1/2}} + \\ + 2 \int_{-\infty}^{\bar{t}} dt' \frac{\ddot{a}_{i} z_{j} + 2 \dot{a}_{i} v_{j} + 2 a_{i} a_{j} + 2 v_{i} \dot{a}_{j} + z_{i} \ddot{a}_{j}}{(\bar{t} - t')^{1/2}}.
\end{multline}
The first two divergent terms in Eq. \eqref{eq:sc_res_int} are cancelled out by the last two terms of the integral \eqref{eq:scal_grav_emit_gen} leaving us with the finite integral.

The derivative of the retarded time $\bar{t}$ is found as
\be
(\partial_{M} \bar{t})^{\rm rad} = \partial_{M} \bar{t} = - \bar{c}_{M}.
\ee
Therefore, the emitted part of the gravitational field due to the scalar field contribution  takes the form
\be
\label{eq:scal_GW_non-rel}
(\partial_{M} \bh_{ij}^{\rm F})^{\rm rad} = - \frac{\mu \kappa_{5} \bar{c}_{M}}{2^{5/2} \pi^{2} r^{3/2}} \int_{-\infty}^{\bar{t}} dt' \frac{\ddot{a}_{i} z_{j} + 2 \dot{a}_{i} v_{j} + 2 a_{i} a_{j} + 2 v_{i} \dot{a}_{j} + z_{i} \ddot{a}_{j}}{(\bar{t} - t')^{1/2}},
\ee
analogous to the emitted part of the gravitational field of point particles \eqref{eq:pp_GW_rad_non-rel_COM}.

%%%%%%%%%%%%%%%%%%%%%%%%%%%%%%%%%%%%%%%%%%%%%%%
\section{Five-dimensional quadrupole formula}\label{V}
%%%%%%%%%%%%%%%%%%%%%%%%%%%%%%%%%%%%%%%%%%%%%%%

Having found the point particles and scalar field contributions into the emitted part of gravitational field, we turn to the calculation of total gravitational radiation power from the binary system.

%%%%%%%%%%%%%%%%%%%%%%%%%%%%%%%%%%%%%%%%%%%%%%%
\subsection{Emitted part of total gravitational field}

Combining the Eqs. \eqref{eq:pp_GW_rad_non-rel_COM} and \eqref{eq:scal_GW_non-rel} we come to the emitted part of the total gravitational field of the binary system
\be
\label{eq:GW_emit_total}
(\partial_{M} \bh_{ij})^{\rm rad} = - \frac{\mu \kappa_{5} \bar{c}_{M}}{2^{5/2} \pi^{2} r^{3/2}} \int_{-\infty}^{\bar{t}} dt' \, \frac{\ddot{a}_{i} z_{j} + 4 \dot{a}_{i} v_{j} + 6 a_{i} a_{j} + 4 v_{i} \dot{a}_{j} + z_{i} \ddot{a}_{j}}{(\bar{t} - t')^{1/2}}.
\ee
Rewriting the denominator in Eq. \eqref{eq:GW_emit_total} as
\be
\ddot{a}_{i} z_{j} + 4 \dot{a}_{i} v_{j} + 6 a_{i} a_{j} + 4 v_{i} \dot{a}_{j} + z_{i} \ddot{a}_{j} = \partial_{t'}^{4} (z_{i} z_{j})
\ee
we arrive at
\be
\label{eq:GW_emit_total_with_deriv}
(\partial_{M} \bh_{ij})^{\rm rad} = - \frac{\mu \kappa_{5} \bar{c}_{M}}{2^{5/2} \pi^{2} r^{3/2}} \int_{-\infty}^{\bar{t}} dt' \, \frac{\partial_{t'}^{4}(z_{i} z_{j})}{(\bar{t} - t')^{1/2}}.
\ee

As we consider the non-relativistic particles, we define, by analogy with the four-dimensional theory \cite{maggiore2008}, the second mass moment of the system as
\be
M^{ij} = \sum_{a} m_{a} \int d^{4}x \, x^{i} x^{j} \, \delta^{(4)}(\mathbf{x} - \mathbf{z}_{a}) = \sum_{a} m_{a} z_{a}^{i} z_{a}^{j}.
\ee
Turning into the center-of-mass frame \eqref{eq:COM_frame_def}, the second mass moment takes the form
\be
\label{eq:sec_mass_mom_CoM}
M_{ij} = \mu z_{i} z_{j},
\ee
where $z^i$ is the relative coordinate of the system defined by Eq. \eqref{eq:rel_coord_def}, and $\mu$ is its reduced mass. Thus, emitted part of the total gravitational field is written as
\be
\label{eq:5D_GW_second_mass_mom}
(\partial_{M} \bh_{ij})^{\rm rad} = - \frac{\kappa_{5} \bar{c}_{M}}{2^{5/2} \pi^{2} r^{3/2}} \int_{-\infty}^{\bar{t}} dt' \, \frac{\ddddot{M}_{ij}}{(\bar{t} - t')^{1/2}}.
\ee
Therefore, by analogy with the gravitational radiation in four dimensions \cite{maggiore2008}, the five-dimensional gravitational radiation of the non-relativistic binary system is determined by its second mass moment.

Moreover, recall that we are interested in the emitted part of gravitational field in the transverse-traceless gauge given by Eq. \eqref{eq:TT_gauge_bh}. Then, taking into account the vanishing traces of the projector \eqref{eq:Lambda_traces} one can replace the second mass moment of the system with its quadrupole moment
\be
\label{eq:5D_quadrupole_def}
Q_{ij} = M_{ij} - \frac{1}{4} \delta_{ij} M_{kk},
\ee
arriving at the emitted part of total gravitational field in the transverse-traceless gauge in form
\begin{align}
\label{eq:GW_TT_quadrupole}
& \left( \partial_{M} h_{ij}^{\rm TT} \right)^{\rm rad} = - \frac{\kappa_{5} \bar{c}_{M}}{2^{5/2} \pi^{2} r^{3/2}} {\cal A}_{ij}^{\rm TT}, \\
\label{eq:curv_A_def}
& {\cal A}_{ij}^{\rm TT} = \int_{-\infty}^{\bar{t}} dt' \frac{\ddddot{Q}_{ij}^{\rm TT}}{(\bar{t} - t')^{1/2}},
\end{align}
where $Q_{ij}^{\rm TT}=\Lambda_{ij,kl}(\mathbf{n})Q_{kl}$. Obtained emitted part of the total gravitational field is analogous to that of the four-dimensional theory \cite{maggiore2008}. However, while in four dimensions gravitational radiation at some moment of time is completely determined by the state of the source at the corresponding retarded moment, in five dimensions it depends on the entire history of the source's motion preceding the retarded time.

%%%%%%%%%%%%%%%%%%%%%%%%%%%%%%%%%%%%%%%%%%%%%%%
\subsection{Quadrupole formula}

Using the Eqs. \eqref{eq:GW_EMT_gen} and \eqref{eq:W_in_D_dimensions} we find the five-dimensional analog of the quadrupole formula \cite{maggiore2008} for the angular distribution of the gravitational radiation power of the non-relativistic binary system
\be
\label{eq:quadr_form_gen}
\frac{dW_{4+1}}{d\Omega_{3}} = \frac{\kappa_{5}}{128 \pi^{4}} \left \langle {\cal A}_{ij}^{\rm TT} {\cal A}_{ij}^{\rm TT} \right \rangle,
\ee
where ${\cal A}_{ij}^{\rm TT}(x)$ is given by the Eq. \eqref{eq:curv_A_def}.

To rewrite Eq. \eqref{eq:quadr_form_gen} in terms of the polarisation amplitudes we follow the derivation analogous to that in four dimensions \cite{maggiore2008}. We note that the contraction of tensors ${\cal A}_{ij}^{\rm TT}$ is invariant under the SO(4)-transformations. Thus, the contraction computed in the arbitrary oriented $x$-frame is equal to the one calculated in the $x'$-frame, where the observation direction vector $\mathbf{n}'$ is aligned with the $x^{\prime 4}$ coordinate
\be
\label{eq:contract_invar}
{\cal A}_{ij}^{\rm TT}(x) {\cal A}_{ij}^{\rm TT}(x) = {\cal A'}_{ij}^{\rm TT}(x') {\cal A'}_{ij}^{\rm TT}(x'),
\ee
see Eq. \eqref{eq:x'_n_vect}. Then, by analogy with the computation of the polarisation amplitudes in Appendix A, we calculate it in the $x^{\prime}$-frame and rewrite the components of tensor ${\cal A'}_{ij}^{\rm TT}$ in terms of ${\cal A}_{ij}^{\rm TT}$ by use of the Eq. \eqref{eq:x_to_x'_tensors}. As a result, using Eq. \eqref{eq:polar_in_x'} we find
\be
{\cal A'}_{ij}^{\rm TT} {\cal A'}_{ij}^{\rm TT} = \tr \left( {\cal A'}^{\rm TT} {\cal A'}^{\rm TT} \right) = 2 \left( {\cal A}_{+}^{2} + {\cal A}_{\times}^{2} + \frac{3}{4} {\cal A}_{\displaystyle \circ}^{2} + {\cal A}_{\oplus}^{2} + {\cal A}_{\otimes}^{2} \right),
\ee
where we have taken into account the definitions of the polarisation amplitudes (\ref{eq:plus_pol_gen} -- \ref{eq:ocross_pol}) and the symmetry of ${\cal A'}_{ij}^{\rm TT}$.

Therefore, the gravitational radiation power takes the following form
\be
\label{eq:5D_GW_rad_polar}
\frac{dW_{4+1}}{d\Omega_{3}} = \frac{\kappa_{5}}{64 \pi^{4}} \left \langle {\cal A}_{+}^{2} + {\cal A}_{\times}^{2} + \frac{3}{4} {\cal A}_{\displaystyle \circ}^{2} + {\cal A}_{\oplus}^{2} + {\cal A}_{\otimes}^{2} \right \rangle.
\ee
The obtained Eq. \eqref{eq:5D_GW_rad_polar} determines the contributions of the independent polarisations into the gravitational radiation of the system. In particular, it is clear that the "breathing" mode ${\cal A}_{\displaystyle \circ}$ carries less energy, than the other polarisation modes of the gravitational field.

%%%%%%%%%%%%%%%%%%%%%%%%%%%%%%%%%%%%%%%%%%%%%%%
\subsection{Binary system in a circular orbit}

As a simple but instructive example, we consider the five-dimensional gravitational radiation from the non-relativistic binary system in a circular orbit. Relative coordinate of such a system is written as
\be
\label{eq:circ_binary_rel_coor}
z^{i} = \lbrace R_{\rm s} \cos{\omega_{\rm s}t}, R_{\rm s} \sin{\omega_{\rm s}t}, 0, 0 \rbrace,
\ee
where $R_{\rm s}$ is the radius of the orbit, and $\omega_{\rm s}$ is the frequency of the orbital motion.

As discussed above, calculating the gravitational radiation one can equivalently use tensor ${\cal A}_{ij}(x)$ defined in terms of the second mass moment \eqref{eq:sec_mass_mom_CoM} or quadrupole moment \eqref{eq:5D_quadrupole_def} of the system. In what follows, we use the former for simplicity
\be
\label{eq:A_ij_sec_mass}
{\cal A}_{ij}(x) = \int_{-\infty}^{\bar{t}} dt' \, \frac{\ddddot{M}_{ij}}{(\bar{t} - t')^{1/2}}.
\ee
The non-vanishing components of its derivative have the form
\begin{align}
\label{eq:sec_mass_mom_non-van_1}
& \ddddot{M}_{11} = - \ddddot{M}_{22} = 8 \mu R_{\rm s}^{2} \omega_{\rm s}^{4} \cos{2\omega_{\rm s} t'}, \\
\label{eq:sec_mass_mom_non-van_2}
& \ddddot{M}_{12} = \ddddot{M}_{21} = 8 \mu R_{\rm s}^{2} \omega_{\rm s}^{4} \sin{2\omega_{\rm s} t'}
\end{align}
From Eqs. \eqref{eq:sec_mass_mom_non-van_1} and \eqref{eq:sec_mass_mom_non-van_2} we find the non-zero components of ${\cal A}_{ij}(x)$
\begin{align}
\label{eq:A_ij_non-van_1}
& {\cal A}_{11}(x) = - {\cal A}_{22}(x) = \sqrt{32 \pi} \mu R_{\rm s}^{2} \omega_{\rm s}^{7/2} \sin\left( 2\omega_{\rm s}\bar{t} + \frac{\pi}{4} \right), \\
\label{eq:A_ij_non-van_2}
& {\cal A}_{12}(x) = {\cal A}_{21}(x) = \sqrt{32 \pi} \mu R_{\rm s}^{2} \omega_{\rm s}^{7/2} \sin\left( 2\omega_{\rm s}\bar{t} - \frac{\pi}{4} \right),
\end{align}
For detailed derivation of the Eqs. \eqref{eq:A_ij_non-van_1} and \eqref{eq:A_ij_non-van_2} see Appendix D.

In accordance with Eq. \eqref{eq:5D_GW_rad_polar}, to compute the power of gravitational radiation of the system we find the polarisations amplitudes of the gravitational field. Substituting into the Eqs. (\ref{eq:plus_pol_gen} - \ref{eq:ocross_pol}) the non-vanishing components of ${\cal A}_{ij}(x)$ tensor \eqref{eq:A_ij_non-van_1} and \eqref{eq:A_ij_non-van_2}, we arrive at
\begin{align}
\label{eq:plus_pol_circ}
& {\cal A}_{+} = - \sqrt{8 \pi} \mu R_{\rm s}^{2} \omega_{\rm s}^{7/2} (1 + \cos^{2}\theta) \sin\left( 2\omega_{\rm s}\bar{t} + \frac{\pi}{4} - 2\phi \right), \\
\label{eq:cross_pol_circ}
& {\cal A}_{\times} = \sqrt{32 \pi} \mu R_{\rm s}^{2} \omega_{\rm s}^{7/2} \cos\theta \cos\left( 2\omega_{\rm s}\bar{t} + \frac{\pi}{4} - 2\phi \right), \\
\label{eq:breath_pol_circ}
& {\cal A}_{\displaystyle \circ} = \frac{\sqrt{32 \pi}}{3} \mu R_{\rm s}^{2} \omega_{\rm s}^{7/2} \left( 2\sin^{2}\theta \cos^{2}\zeta - \cos^{2}\theta + 1 \right) \sin\left( 2\omega_{\rm s}\bar{t} + \frac{\pi}{4} - 2\phi \right), \\
\label{eq:oplus_pol_circ}
& {\cal A}_{\oplus} = \sqrt{32 \pi} \mu R_{\rm s}^{2} \omega_{\rm s}^{7/2} \sin\theta \cos\zeta \cos\left( 2\omega_{\rm s}\bar{t} + \frac{\pi}{4} - 2\phi \right), \\
\label{eq:otimes_pol_circ}
& {\cal A}_{\otimes} = \sqrt{8 \pi} \mu R_{\rm s}^{2} \omega_{\rm s}^{7/2} \sin{2\theta} \cos\zeta \sin\left( 2\omega_{\rm s}\bar{t} + \frac{\pi}{4} - 2\phi \right).
\end{align}
Note that for the observation point lying on the brane $\zeta = \pi/2$ the "breathing" mode \eqref{eq:breath_pol_circ} is non-zero, while the ${\cal A}_{\oplus}$ and ${\cal A}_{\otimes}$ polarisations vanish here and, therefore, observer on the brane detect only three independent polarisations of the gravitational waves, as was found earlier in Ref. \cite{Andriot2017}.

Averaging the squares of polarisation amplitudes in Eq. \eqref{eq:5D_GW_rad_polar} over the period of orbital motion  
\be
\left \langle \sin^{2}\left( 2\omega_{\rm s}\bar{t} + \frac{\pi}{4} - 2\phi \right) \right \rangle = \left \langle \cos^{2}\left( 2\omega_{\rm s}\bar{t} + \frac{\pi}{4} - 2\phi \right) \right \rangle = \frac{1}{2},
\ee
we come to the angular distribution of the gravitational radiation power of the binary system in  circular motion
\begin{equation}
\label{eq:5D_GW_ang_dist}
\frac{dW_{4+1}^{\rm circ}}{d\Omega_{3}} = \frac{\kappa_{5} \mu^{2} R_{\rm s}^{4} \omega_{\rm s}^{7}}{4 \pi^{3}} \left \lbrack \left( \frac{1 + \cos^{2}\theta}{2} \right)^{2} + \cos^{2}\theta + \sin^{4}\theta \frac{( 1 + 2 \cos^{2}\zeta )^{2}}{12} + \sin^{2}\theta \cos^{2}\zeta ( 1 + \cos^{2}\theta ) \right \rbrack.
\end{equation}
Integrating the Eq. \eqref{eq:5D_GW_ang_dist}, we obtain the total gravitational radiation power of the non-relativistic binary system on the circular orbit
\be
\label{eq:5D_GW_tot_power_circ}
W_{4+1}^{\rm circ} = \frac{5}{9\pi} \kappa_{5} \mu^{2} R_{\rm s}^{4} \omega_{\rm s}^{7}.
\ee

%%%%%%%%%%%%%%%%%%%%%%%%%%%%%%%%%%%%%%%%%%%%%%%
\subsection{Evolution of the orbit}

As we emphasized from the very beginning, this calculation should be considered as a mathematically consistent but physically unrealistic model, so we refrain from making any quantitative physical predictions. In particular, the evolution of the orbit, which we consider in this subsection as a logical extension of our model, cannot be considered as an attempt to extract experimentally available predictions and gives a result that clearly contradicts the observational data of binary pulsars. However, we present this calculation simply as an example of a toy model showing how different frequency dependences give different predictions for the lifetime of a binary system.

Consider the orbital evolution of the system  due to gravitational radiation.  As in the non-relativistic limit the gravitational-wave energy loss over the period of system's motion is small compared to its total energy, we assume that the system moves on a quasi-circular orbit. Indeed, by analogy with the four-dimensional theory \cite{maggiore2008}, we consider the orbital radius $R_{\rm s}(t)$ and the frequency of orbital motion $\omega_{\rm s}(t)$ as the slowly varying functions of time satisfying condition
\be
\label{eq:quasicirc_def}
-\dot{R}_{\rm s} \ll \omega_{\rm s} R_{\rm s},
\ee
corresponding to the slow shrinking of orbit. The orbital evolution is governed by the energy conservation law
\be
\label{eq:en_bal_eq}
\frac{dE_{\rm tot}}{dt} = - W_{4+1}^{\rm circ},
\ee
where $E_{\rm tot}$ is the total mechanical energy of the non-relativistic binary system, and gravitational radiation power is given by the Eq. \eqref{eq:5D_GW_tot_power_circ}.

We rewrite the mechanical energy of the system and its radiation power in terms of the frequency of orbital motion by use of its relation to the orbital radius. Substituting into the Eq. \eqref{eq:pp_EOM_non-rel} relative coordinate of the system \eqref{eq:circ_binary_rel_coor}, up to the leading order contributions, we arrive at
\be
\label{eq:freq_radius_rel}
\omega_{\rm s}^{2} = \frac{g_1 g_2}{\mu R_{\rm s}^{3}}.
\ee
Using the Eq. \eqref{eq:freq_radius_rel} we rewrite the energy of the system and its gravitational radiation power as
\begin{align}
\label{eq:tot_en_freq}
& E_{\rm tot} = \frac{\mu \mathbf{v}^{2}}{2} - \frac{g_{1} g_{2}}{|\mathbf{z}|} = - \frac{1}{2} (\sqrt{\mu} g_1 g_2)^{2/3} \omega_{\rm s}^{2/3}, \\
\label{eq:grav_power_freq}
& W_{4+1}^{\rm circ} = \frac{5}{9\pi} \kappa_{5} (\sqrt{\mu} g_1 g_2)^{4/3} \omega_{\rm s}^{13/3}.
\end{align}
Therefore, substituting the Eqs. \eqref{eq:tot_en_freq} and \eqref{eq:grav_power_freq} into the energy conservation law \eqref{eq:en_bal_eq}, we come to the equation for the evolution of the orbital frequency
\begin{align}
\label{eq:or_freq_ev}
&\dot{\omega}_{\rm s} = \frac{5 \kappa_{5}}{3\pi} (\sqrt{\mu} g_1 g_2)^{2/3} \omega_{\rm s}^{14/3},\\
\label{eq:qua_cir_freq}
&\dot{\omega}_{\rm s} \ll \omega_{\rm s}^{2},
\end{align}
where Eq. \eqref{eq:qua_cir_freq} is the condition for the quasi-circular motion \eqref{eq:quasicirc_def} in terms of the orbital frequency.

Solution to the Eq. \eqref{eq:or_freq_ev} is found as
\be
\label{eq:freq_evol}
\omega_{\rm s}(t) = \left( \frac{9\pi}{55\kappa_{5}} \right)^{3/11} (\sqrt{\mu} g_1 g_2)^{-2/11} s^{-3/11},
\ee
where $s=t_{\rm coal} - t$, and $t_{\rm coal}$ is the moment of coalescence of the binary system. From the Eq. \eqref{eq:freq_radius_rel} we find the equation for the orbital radius
\be
\label{eq:orbit_shrink}
\dot{R}_{\rm s} = - \frac{2}{3} R_{\rm s} \frac{\dot{\omega}_{\rm s}}{\omega_{\rm s}},
\ee
and from the Eq. \eqref{eq:freq_evol} we determine the evolution of the orbital radius as
\be
\label{eq:orb_rad_evol}
R_{\rm s}(t) = R_{0} \left( \frac{t_{\rm coal} - t}{t_{\rm coal} - t_{0}} \right)^{2/11},
\ee
where $R_{0}$ is the orbital radius at some moment of time $t_{0}$. Note that the Eqs. \eqref{eq:freq_evol} and \eqref{eq:orb_rad_evol} differ significantly from that in four dimensions \cite{maggiore2008}
\be
\omega_{\rm s,\,4D}(t) \sim (t_{\rm coal} - t)^{-3/8}, \quad R_{\rm s,\,4D}(t) \sim (t_{\rm coal} - t)^{1/4},
\ee
as could be expected, given that in our work gravity is described by the five-dimensional GR with the infinite-volume extra dimension being phenomenologically non-viable.

%%%%%%%%%%%%%%%%%%%%%%%%%%%%%%%%%%%%%%%%%%%%%%%
\section{Conclusion}\label{VI}
%%%%%%%%%%%%%%%%%%%%%%%%%%%%%%%%%%%%%%%%%%%%%%%

In this article, we have considered (partially inspired by the DGP model) a mixed environment with five-dimensional gravity and four-dimensional matter consisting of point particles and a scalar field. The mechanism of dynamic confinement was not imposed, and the localization of particle motion in the brane was assumed to be purely kinematic due to the appropriate choice of initial conditions. The physical viability of this setup as a cosmological model is ruled out, but mathematically it is a consistent model that opens the way for calculating bulk gravitational radiation in the framework of linearized gravity. This model reflects a typical feature of radiation in odd space-time dimensions, which is the appearance of tails due to violation of the Huygens principle. 
 
We have introduced techniques that can be useful in other problems of radiation in odd-dimensional spacetime, namely, the extraction of the radiative part of the retarded potential using a modification of the Rohrlich-Teitelboim approach. It has also been shown that the contribution of field stresses through which two point masses interact to gravitational radiation can be explicitly taken into account using the DIRE approach to post-Newtonian expansions in four dimensions. As a result, it turned out that in odd dimensions the radiated part of the gravitational field depends on the entire history of the system's motion preceding the retarded time, in contrast to even dimensions, where it depends only on the state of the system at the retarded moment of proper time. Another interesting feature of odd dimensions is that the retarded gravitational field is given by the sum of separately divergent integrals. However, these divergences cancel each other out, and the resulting retarded field and its radiated part are finite. 

We considered the gravitational radiation from the binary system of two non-relativistic point particles  deriving the five-dimensional analog of the quadrupole formula for   gravitational radiation power \eqref{eq:quadr_form_gen}. We found that in contrast with the ultrarelativistic synchrotron radiation in odd dimensions \cite{Galtsov:2020hhn}, the dependence of the gravitational radiation of a nonrelativistic source on its motion history is not localized near the retarded time. 

Another feature concerns polarization. In five dimensions free gravitational waves have five polarizations. If motion is restricted to four-dimensional branes, an observer living on the brane will see  three of them -- two standard "cross" and "plus" polarisations and an additional "breathing" mode. This feature was also provided from other \cite{Andriot2017} considerations and may be of interest for more realistic models  

Now let's briefly discuss the relationship of our results with other work. Our quadrupole formula agrees with the formula obtained by Chu \cite{Chu:2021uea} using the Fourier transforms of the retarded Green's functions. In addition, our results fit into the general schemes for the quadrupole formula and the radiation power of the system in a circular orbit obtained by Cardoso {\it et al.} \cite{Cardoso:2002pa} for the arbitrary even dimensions. We also expect that the resulting formula for the radiation power of the system at the circular orbit can be verified using the results obtained by Cardoso {\it et al.} \cite{Cardoso:2008gn} in effective field theory approach framework.

\section*{Acknowledgements}

The work of M. Kh. was supported by the “BASIS” Foundation Grant No. 20-2-10-8-1. This research was also supported by the Russian Foundation for Basic Research on the project 20-52-18012Bulg-a, and the Scientific and Educational School of Moscow State University “Fundamental and Applied Space Research”.

\appendix

%%%%%%%%%%%%%%%%%%%%%%%%%%%%%%%%%%%%%%%%%%%%%%%
\section{Polarizations in an arbitrary observation direction}
%%%%%%%%%%%%%%%%%%%%%%%%%%%%%%%%%%%%%%%%%%%%%%%

The following derivation is analogous to that in \cite{maggiore2008}. We calculate the polarizations of some symmetric rank-2 tensor ${\cal A}_{ij}^{\rm TT}$ with respect to the appropriately chosen frame. Then, we rewrite its components used in the polarizations in terms of the components defined with respect to the arbitrarily oriented frame.

We begin with the tensor ${\cal A}_{ij}^{\prime}$ in the $x'$-frame, where the observation direction vector $\mathbf{n}^{\prime}$ is aligned with the $x^{\prime 4}$-coordinate
\be
\label{eq:x'_n_vect}
{n}^{\prime i} = \lbrace 0, 0, 0, 1\rbrace.
\ee
Using the Eq. \eqref{eq:TT_gauge_def} we find the ${\cal A}_{ij}^{\prime}$ tensor in the transverse-traceless gauge as
\be
\label{eq:polar_in_x'}
{{\cal A}'}_{ij}^{\rm TT} = 
\begin{pmatrix}
{\cal A}_{+} - \displaystyle \frac{1}{2} {\cal A}_{\displaystyle \circ} & {\cal A}_{\times} & {\cal A}_{\oplus} & 0 \\
{\cal A}_{\times} & - {\cal A}_{+} - \displaystyle \frac{1}{2} {\cal A}_{\displaystyle \circ} & {\cal A}_{\otimes} & 0 \\
{\cal A}_{\oplus} & {\cal A}_{\otimes} & {\cal A}_{\displaystyle \circ} & 0 \\
0 & 0 & 0 & 0
\end{pmatrix},
\ee
where the polarizations are defined by analogy with the Eqs. \eqref{eq:pl_cr_pol_part} and \eqref{eq:breath_pol_part}
\begin{align}
\label{eq:x_prime_pol_1}
&{\cal A}_{+} = \frac{1}{2} \left( {\cal A}_{11}^{\prime} - {\cal A}_{22}^{\prime} \right), \quad {\cal A}_{\times} = {\cal A}_{12}^{\prime}, \quad {\cal A}_{\displaystyle \circ} = \frac{2}{3} {\cal A}_{33}^{\prime} - \frac{1}{3} \left( {\cal A}_{11}^{\prime} + {\cal A}_{22}^{\prime} \right), \\
\label{eq:x_prime_pol_2}
&{\cal A}_{\oplus} = {\cal A}_{13}^{\prime}, \quad {\cal A}_{\otimes} = {\cal A}_{23}^{\prime}.
\end{align}

Now we rewrite the components of ${\cal A}_{ij}^{\prime}$ in terms of the components of ${\cal A}_{ij}$ defined with respect to the arbitrary oriented $x$-frame, where the observation direction vector $\mathbf{n}$ has the form
\be
{n}^{i} = \lbrace \cos\phi \sin\theta \sin\zeta, \sin\phi \sin\theta \sin\zeta, \cos\theta \sin\zeta, \cos\zeta \rbrace,
\ee
and we use the hyperspherical coordinates \cite{Landim:2009arx}. It is related to $\mathbf{n}^{\prime}$ by the SO(4)-transformation
\be
n^{i} = R^{i}_{\, j} n^{\prime \, j},
\ee
where the transformation matrix $R^{i}_{\, j}$ is written as
\be
R^{i}_{\, j} = 
\begin{pmatrix}
\sin\phi & \cos\phi \cos\theta & \cos\phi \sin\theta \cos\zeta & \cos\phi \sin\theta \sin\zeta \\
-\cos\phi & \sin\phi \cos\theta & \sin\phi \sin\theta \cos\zeta & \sin\phi \sin\theta \sin\zeta \\
0 & -\sin\theta & \cos\theta \cos\zeta & \cos\theta \sin\zeta \\
0 & 0 & -\sin\zeta & \cos\zeta
\end{pmatrix}.
\ee
Analogously, components of ${\cal A}_{ij}^{\prime}$ are written in terms of the components of ${\cal A}_{ij}$ as
\be
\label{eq:x_to_x'_tensors}
{\cal A}_{mn}^{\prime} = \left( R^{\rm T} {\cal A} R \right)_{mn},
\ee
where $(PQ)_{ij} = P_{ik}Q_{kj}$ is the matrix product and $R^{\rm T}$ is the transposed transformation matrix.

As a result, after some algebra we obtain the polarizations for the arbitrary observation direction $\mathbf{n}$. First, the standard "plus" and "cross" polarizations take the form
\begin{align}
{\cal A}_{+} & = \frac{1}{2} \left \lbrack {\cal A}_{11} ( \sin^{2}\phi - \cos^{2}\phi \cos^{2}\theta ) + {\cal A}_{22} ( \cos^{2}\phi - \sin^{2}\phi \cos^{2}\theta ) - A_{33} \sin^{2}\theta - \right. \nonumber \\
\label{eq:plus_pol_gen}
& - \left. {\cal A}_{12} ( 1 + \cos^{2}\theta ) \sin{2\phi} + ( {\cal A}_{13} \cos\phi + {\cal A}_{23} \sin\phi ) \sin{2\theta} \right \rbrack, \\
\label{eq:cross_pol_gen}
{\cal A}_{\times} & = \frac{1}{2} ( {\cal A}_{11} - {\cal A}_{22} ) \sin{2\phi} \cos\theta - {\cal A}_{12} \cos{2\phi} \cos\theta - ( {\cal A}_{13} \sin\phi - {\cal A}_{23} \cos\phi ) \sin\theta.
\end{align}
Note that the obtained polarisations \eqref{eq:plus_pol_gen} and \eqref{eq:cross_pol_gen} are that of the four-dimensional theory \cite{maggiore2008}. "Breathing" mode is found as
\begin{multline}
\label{eq:breath_pol_gen}
{\cal A}_{\displaystyle \circ} = \frac{1}{3} \left \lbrace {\cal A}_{11} \lbrack ( 2 \cos^{2}\zeta + 1 ) \cos^{2}\phi \sin^{2}\theta - 1 \rbrack + {\cal A}_{22} \lbrack ( 2 \cos^{2}\zeta + 1 ) \sin^{2}\phi \sin^{2}\theta - 1 \rbrack + {\cal A}_{33} \right. \times \\ \times \lbrack ( 2 \cos^{2}\zeta + 1 ) \cos^{2}\theta - 1 \rbrack + 2 {\cal A}_{44} \sin^{2}\zeta + \lbrack {\cal A}_{12} \sin{2\phi} \sin^{2}\theta + ( {\cal A}_{13} \cos\phi + {\cal A}_{23} \sin\phi ) \times \\ \times \left. \sin{2\theta} \rbrack ( 2 \cos^{2}\zeta + 1 ) - 2 \lbrack ( {\cal A}_{14} \cos\phi + {\cal A}_{24} \sin\phi ) \sin\theta + {\cal A}_{34} \cos\theta \rbrack \sin{2\zeta} \right \rbrace.
\end{multline}
Note that it is non-vanishing even when the observation point and the source are on the brane $\zeta = \pi/2, \, {\cal A}_{i4}=0$. Remaining polarizations ${\cal A}_{\oplus}$ and ${\cal A}_{\otimes}$ take the form
\begin{align}
{\cal A}_{\oplus} & = \frac{1}{2} \left( {\cal A}_{11} - {\cal A}_{22} \right) \sin{2\phi} \sin\theta \cos\zeta - {\cal A}_{12} \cos{2\phi} \sin\theta \cos\zeta + \left( {\cal A}_{13} \sin\phi - {\cal A}_{23} \cos\phi \right) \times \nonumber \\
\label{eq:oplus_pol}
& \times \cos\theta \cos\zeta - \left( {\cal A}_{14} \sin\phi - {\cal A}_{24} \cos\phi \right) \sin\zeta, \\
{\cal A}_{\otimes} & = \frac{1}{2} \left \lbrace \left( {\cal A}_{11} \cos^{2}\phi + {\cal A}_{22} \sin^{2}\phi - {\cal A}_{33} \right) \sin{2\theta} \cos\zeta + {\cal A}_{12} \sin{2\phi} \sin{2\theta} \cos\zeta + 2 \left( {\cal A}_{13} \cos\phi + \right. \right. \nonumber \\
\label{eq:ocross_pol}
& + \left. \left. {\cal A}_{23} \sin\phi \right) \cos{2\theta} \cos\zeta - 2 \left( {\cal A}_{14} \cos\phi + {\cal A}_{24} \sin\phi \right) \cos\theta \sin\zeta + {\cal A}_{34} \sin\theta \sin\zeta \right \rbrace.
\end{align}
In contrast with the "breathing" mode, they vanish when the observation point and the source are on the brane.

%%%%%%%%%%%%%%%%%%%%%%%%%%%%%%%%%%%%%%%%%%%%%%%
\section{Regularisation of the point particles term}

Consider the first two terms in Eq. \eqref{eq:pp_GW_non-rel_non-int} and split it into three integrals
\begin{align}
\label{eq:I_i_def_1}
& I_{ij} = \int_{-\infty}^{\bar{t}} dt' \left \lbrack \frac{15}{4} v_{i} v_{j} \frac{\mathbf{n}\mathbf{s}(t') - \mathbf{n}\bar{\mathbf{v}}}{(\bar{t} - t')^{5/2}} + \frac{3}{2} \frac{v_{i} v_{j}}{(\bar{t} - t')^{5/2}} \right \rbrack = I_{ij}^{1} + I_{ij}^{2} + I_{ij}^{3}, \\
& I_{ij}^{1} = \frac{15}{4} \int_{-\infty}^{\bar{t}} dt' \, v_{i} v_{j} \frac{\mathbf{n}\mathbf{s}(t')}{(\bar{t} - t')^{5/2}}, \quad I_{ij}^{2} = - \frac{15}{4} \int_{-\infty}^{\bar{t}} dt' \, v_{i} v_{j} \frac{\mathbf{n}\bar{\mathbf{v}}}{(\bar{t} - t')^{5/2}}, \\
\label{eq:I_i_def_3}
& I_{ij}^{3} = \frac{3}{2} \int_{-\infty}^{\bar{t}} dt' \, \frac{v_{i} v_{j}}{(\bar{t} - t')^{5/2}}.
\end{align}
By analogy with the regularisation of scalar field \eqref{eq:5D_sc_lead_term_exp}, we introduce the regularising parameter $\epsilon \to +0$ into the upper integration limit $\bar{t} \to \bar{t} - \epsilon$ and integrate Eqs. (\ref{eq:I_i_def_1}--\ref{eq:I_i_def_3}) by parts reducing the powers on denominators to $1/2$.

Using the Eq. \eqref{eq:s_non-rel} and integrating the $I_{ij}^{1}$ three times by parts we extract its divergent contributions coming from the upper integration limit
\begin{multline}
\label{eq:div_I_1}
I_{ij}^{1} = \frac{5}{2} \lim_{\epsilon \to +0} \bar{v}_{i} \bar{v}_{j} \frac{\mathbf{n}\bar{\mathbf{v}}}{\epsilon^{3/2}} - 2 \lim_{\epsilon \to +0} \bar{v}_{i} \bar{v}_{j} \frac{\mathbf{n}\bar{\mathbf{a}}}{\epsilon^{1/2}} - 10 \lim_{\epsilon \to +0} \bar{a}_{(i} \bar{v}_{j)} \frac{\mathbf{n}\bar{\mathbf{v}}}{\epsilon^{1/2}} + 2 \int_{-\infty}^{\bar{t}} dt' \, v_{i} v_{j} \frac{\mathbf{n}\dot{\mathbf{a}}}{(\bar{t} - t')^{1/2}} + \\ + 12 \int_{-\infty}^{\bar{t}} dt' \, a_{(i} v_{j)} \frac{\mathbf{n}\mathbf{a}}{(\bar{t} - t')^{1/2}} + 6 \int_{-\infty}^{\bar{t}} dt' \left( \dot{a}_{i} v_{j} + 2 a_{i} a_{j} + v_{i} \dot{a}_{j} \right) \frac{\mathbf{n}\mathbf{v}}{(\bar{t} - t')^{1/2}} - \\ - 2 \int_{-\infty}^{\bar{t}} dt' \left( \ddot{a}_{i} v_{j} + 3 \dot{a}_{i} a_{j} + 3 a_{i} \dot{a}_{j} + v_{i} \ddot{a}_{j} \right) \frac{\mathbf{n}\bar{\mathbf{z}} - \mathbf{n}\mathbf{z}}{(\bar{t} - t')^{1/2}},
\end{multline}
where we have taken into account that $\mathbf{s}(t) \to \bar{\mathbf{v}}, \, t \to \bar{t}$. After analogous transformations, the $I_{ij}^{2}$ integral is found as
\be
\label{eq:div_I_2}
I_{ij}^{2} = - \frac{5}{2} \lim_{\epsilon \to +0} \bar{v}_{i} \bar{v}_{j} \frac{\mathbf{n}\bar{\mathbf{v}}}{\epsilon^{3/2}} + 10 \lim_{\epsilon \to +0} \bar{a}_{(i} \bar{v}_{j)} \frac{\mathbf{n}\bar{\mathbf{v}}}{\epsilon^{1/2}} - 5 \int_{-\infty}^{\bar{t}} dt' \left( \dot{a}_{i} v_{j} + 2 a_{i} a_{j} + v_{i} \dot{a}_{j} \right) \frac{\mathbf{n}\bar{\mathbf{v}}}{(\bar{t} - t')^{1/2}}.
\ee
The first two terms of Eq. \eqref{eq:div_I_2} cancel out the two divergent terms in the Eq. \eqref{eq:div_I_1}. The last integral $I_{ij}^{3}$ is transformed to the following form
\be
\label{eq:div_I_3}
I_{ij}^{3} = \lim_{\epsilon \to +0} \frac{\bar{v}_{i} \bar{v}_{j}}{\epsilon^{3/2}} - 4 \lim_{\epsilon \to +0} \frac{\bar{a}_{(i} \bar{v}_{j)}}{\epsilon^{1/2}} + 2 \int_{-\infty}^{\bar{t}} dt' \frac{\dot{a}_{i} v_{j} + 2 a_{i} a_{j} + v_{i} \dot{a}_{j}}{(\bar{t} - t')^{1/2}}.
\ee

The remaining divergent terms in Eqs. \eqref{eq:div_I_2} and \eqref{eq:div_I_3} are cancelled out by the "counter-terms" contained in the Eq. \eqref{eq:pp_GW_non-rel_non-int}
\begin{align}
& - \frac{3}{2} \lim_{\epsilon \to +0} \int_{-\infty}^{\bar{t}-\epsilon} dt' \frac{\bar{v}_{i} \bar{v}_{j}}{(\bar{t} - t')^{5/2}} = - \lim_{\epsilon \to +0} \frac{\bar{v}_{i} \bar{v}_{j}}{\epsilon^{3/2}}, \\
& \lim_{\epsilon \to +0} \int_{-\infty}^{\bar{t}-\epsilon} dt' \frac{2 \bar{a}_{(i} \bar{v}_{j)} + \bar{v}_{i} \bar{v}_{j} \mathbf{n}\bar{\mathbf{a}}}{(\bar{t} - t')^{3/2}} = 2 \lim_{\epsilon \to +0} \frac{2 \bar{a}_{(i} \bar{v}_{j)} + \bar{v}_{i} \bar{v}_{j} \mathbf{n}\bar{\mathbf{a}}}{\epsilon^{1/2}},
\end{align}
leaving us with the convergent integral \eqref{eq:pp_GW_non-rel_mixed} for the emitted part of gravitational derivative.

%%%%%%%%%%%%%%%%%%%%%%%%%%%%%%%%%%%%%%%%%%%%%%%
\section{Spatial integral in scalar field contribution}

Spatial integral arising in the Eqs. \eqref{eq:tail_int_separated} and \eqref{eq:cone_int_lead_multipole} has the form
\be
\label{eq:sp_int_def}
\int_{{\cal R}} d^{3}x' \, \varphi_{1}(x') \partial_{i}^{\, \prime} \partial_{j}^{\, \prime} \varphi_{2}(x'),
\ee
where the subscript denotes the integration over the interior of the two-dimensional sphere of radius $\cal R$ with center at the origin.

Given the non-relativistic motion of particles and neglecting the retardation of the fields in the near zone, the scalar fields are given by the Eq. \eqref{eq:ret_sc_non-rel}. Rewriting the particles' world lines in terms of the relative coordinate \eqref{eq:rel_coord_def} we obtain
\be
\int_{\cal R} d^{3}x' \frac{g_1}{|\vec{x}' + s_2 \vec{z}(t')|} \partial_{i}^{\,\prime} \partial_{j}^{\,\prime} \frac{g_{2}}{|\vec{x}' - s_1 \vec{z}(t')|},
\ee
where $s_1 = m_1/M$, and $s_2=m_2/M$. We shift the origin as
\be
\vec{y} = \vec{x}' - s_1 \vec{z},
\ee
to get under derivatives the inverse distance from the origin. We, also, change the integration region to the interior of the sphere centered at $\vec{y}=0$. New integration region deviates from the original one just in the regions $|\vec{y}| \sim {\cal R}$, where the energy-momentum density of the scalar field is vanishingly small. Therefore, we neglect these deviations. As a result, we arrive at the spatial integral in form
\be
\label{eq:spat_int_origin_shift}
3 g_1 g_2 \delta_{i\mathbf{i}} \delta_{j\mathbf{j}} \int dy \, d\Omega_{2} \frac{\hat{n}^{\mathbf{i}} \hat{n}^{\mathbf{j}}}{y|\vec{y} + \vec{z}|}, \quad y=|\vec{y}|, \quad \hat{n}^{\mathbf{i}} = \frac{y^{\mathbf{i}}}{y},
\ee
where the bold indices denote the coordinates on the brane $\mathbf{i},\mathbf{j}=\overline{1,3}$, we use the spherical coordinates, and $\Omega_{2}$ is an angular element on the sphere.

To compute the integral over the sphere in Eq. \eqref{eq:spat_int_origin_shift} we expand the integrand into the spherical harmonics (see, e.g., \cite{Pati:2000vt,Will:1996zj,Thorne:1980ru}) as
\be
\frac{1}{|\vec{y} + \vec{z}|} = \sum_{l,m} \frac{4\pi(-1)^{l}}{2l+1} \frac{r_{<}^{l}}{r_{>}^{l+1}} Y_{lm}^{*} (n_{z}^{\mathbf{i}}) Y_{lm}(\hat{n}^{\mathbf{i}}),
\ee
where $r_{>(<)}$ is the greater (lesser) of $y$ and $z=|\vec{z}|$. Then, the integral in Eq. \eqref{eq:spat_int_origin_shift} splits into two terms corresponding to the integration over the interior of the sphere of radius $z$ and over the spherical shell $y \in (z,{\cal R})$
\begin{multline}
\label{eq:spat_int_split}
\int dy \, d\Omega_{2} \frac{\hat{n}^{\mathbf{i}} \hat{n}^{\mathbf{j}}}{y|\vec{y} + \vec{z}|} = 4 \pi \sum_{l,m} \int_{0}^{z} dy \, \frac{y^{l-1}}{z^{l+1}} \int d\Omega_{2} \, \frac{(-1)^{l}}{2l+1} \left( \hat{n}^{\mathbf{i}} \hat{n}^{\mathbf{j}} - \frac{1}{3} \delta^{\mathbf{i}\mathbf{j}} \right) Y_{lm}^{*} (n_{z}^{\mathbf{i}}) Y_{lm}(\hat{n}^{\mathbf{i}}) + \\ + 4 \pi \sum_{l,m} \int_{z}^{\cal R} dy \, \frac{z^{l}}{y^{l+2}} \int d\Omega_{2} \, \frac{(-1)^{l}}{2l+1} \left( \hat{n}^{\mathbf{i}} \hat{n}^{\mathbf{j}} - \frac{1}{3} \delta^{\mathbf{i}\mathbf{j}} \right) Y_{lm}^{*} (n_{z}^{\mathbf{i}}) Y_{lm}(\hat{n}^{\mathbf{i}}),
\end{multline}
where we added the Kronecker deltas to the products of two unit vectors, given that $\delta_{i\mathbf{i}} \delta_{j\mathbf{j}} \delta^{\mathbf{i}\mathbf{j}} = \delta_{ij}$ and that in the transverse-traceless gauge these terms vanish.

The angular integral in the first term of Eq. \eqref{eq:spat_int_split} is computed by use of the formulas of Ref. \cite{Pati:2000vt,Thorne:1980ru,Will:1996zj}
\be
\label{eq:spher_har_int}
\sum_{m} \int d\Omega_{2} \left( \hat{n}^{\mathbf{i}} \hat{n}^{\mathbf{j}} - \frac{1}{3} \delta^{\mathbf{i}\mathbf{j}} \right) Y_{lm}^{*} (n_{z}^{\mathbf{i}}) Y_{lm}(\hat{n}^{\mathbf{i}}) = \left( \hat{n}^{\mathbf{i}}_{z} \hat{n}^{\mathbf{j}}_{z} - \frac{1}{3} \delta^{\mathbf{i}\mathbf{j}} \right) \delta_{l2}.
\ee
The remaining radial integral in the first term of Eq. \eqref{eq:spat_int_split} is found as
\be
\label{eq:spat_int_inter}
\frac{4\pi}{5} \left( \hat{n}^{\mathbf{i}}_{z} \hat{n}^{\mathbf{j}}_{z} - \frac{1}{3} \delta^{\mathbf{i}\mathbf{j}} \right) \int_{0}^{z} dy \, \frac{y}{z^3} = \frac{2\pi}{5} \frac{z^{\mathbf{i}} z^{\mathbf{j}}}{z^{3}},
\ee
where we omitted the Kronecker delta term in the right-hand side. Using the Eq. \eqref{eq:spher_har_int}, we compute the second term in Eq. \eqref{eq:spat_int_split} arriving at
\be
\label{eq:spat_int_exter}
- \frac{4\pi}{15} \frac{z^{\mathbf{i}} z^{\mathbf{j}}}{{\cal R}^{3}} + \frac{4\pi}{15} \frac{z^{\mathbf{i}} z^{\mathbf{j}}}{z^{3}}.
\ee
Omitting, as discussed above, the term proportional to the inverse near zone radius and combining the Eqs. \eqref{eq:spat_int_inter} and \eqref{eq:spat_int_exter} we obtain the spatial integral \eqref{eq:sp_int_def} as
\be
\int_{{\cal R}} d^{3}x' \, \varphi_{1}(x') \partial_{i}^{\, \prime} \partial_{j}^{\,\prime} \varphi_{2}(x') = 2\pi g_1 g_2 \frac{z^{i} z^{j}}{z^{3}},
\ee

Finally, using the particles equation of motion we rewrite it as
\be
\int_{{\cal R}} d^{3}x^{\prime} \, \varphi_{1}(x') \partial_{i}^{\,\prime} \partial_{j}^{\,\prime} \varphi_{2}(x') = - \pi \mu \lbrack a^{i} z^{j} + z^{i} a^{j} \rbrack.
\ee

%%%%%%%%%%%%%%%%%%%%%%%%%%%%%%%%%%%%%%%%%%%%%%%
\section{Radiation of circular binary system}

Here, we calculate only the ${\cal A}_{11}(x)$ component, and the remaining ones are computed in analogous way. Substituting into the Eq. \eqref{eq:A_ij_sec_mass} the derivative of the second mass moment \eqref{eq:sec_mass_mom_non-van_1} we arrive at
\be
{\cal A}_{11}(x) = 8 \mu R_{\rm s}^{2} \omega_{\rm s}^{4} \int_{-\infty}^{\bar{t}} dt' \frac{\cos{2\omega_{\rm s} t}}{(\bar{t} - t')^{1/2}}.
\ee
Introducing new integration variable $s = \bar{t} - t'$ and expanding the cosine of difference, we obtain
\be
{\cal A}_{11}(x) = 8 \mu R_{\rm s}^{2} \omega_{\rm s}^{4} \int_{0}^{+\infty} \frac{ds}{\sqrt{s}} \left( \cos{2\omega_{\rm s}\bar{t}} \cos{2\omega_{\rm s}s} + \sin{2\omega_{\rm s}\bar{t}} \sin{2\omega_{\rm s}s} \right).
\ee
Two remaining integrals are just the Fresnel integrals \cite{zwillinger2014table}
\be
\int_{0}^{+\infty} dx \frac{\cos(ax)}{\sqrt{x}} = \int_{0}^{+\infty} dx \frac{\sin(ax)}{\sqrt{x}} = \sqrt{\frac{\pi}{2a}}, \, a>0.
\ee
Therefore, using the formula for the sum of sine and cosine we find
\be
{\cal A}_{11}(x) = \sqrt{32\pi} \mu R_{\rm s}^{2} \omega_{\rm s}^{7/2} \sin\left( 2\omega_{\rm s}\bar{t} + \frac{\pi}{4} \right).
\ee

%%%%%%%%%%%%%%%%%%%%%%%%%%%%%%%%%%%%%%%%%%%%%%%
%\bibliographystyle{plain}
%\bibliography{Bibliography}
%%%%%%%%%%%%%%%%%%%%%%%%%%%%%%%%%%%%%%%%%%%%%%%

\end{document}